\documentclass[prstab,twocolumn,showpacs,preprintnumbers,amsmath,amssymb]{revtex4}
\usepackage[english]{babel}
\usepackage{indentfirst}
\usepackage{epsfig}
\usepackage{amsfonts,amssymb,latexsym,amsmath,enumerate}
\usepackage{bm}
\def\curl{{\rm curl}}
\def\div{{\rm div}}
\def\grad{{\rm grad}}

\begin{document}


\title{Universal description for different types of polarization radiation}
\author{D.V. Karlovets}
\email{karlovets@tpu.ru}
\author{A.P. Potylitsyn}
\affiliation{Tomsk Polytechnic University, Lenina ave. 30, Tomsk 634050, Russian Federation}

\date{\today}

\begin{abstract}
When a charged particle moves nearby a spatially inhomogeneous condensed medium or inside it, different types of radiation may arise: 
Diffraction radiation (DR), Smith-Purcell radiation (SPR), Transition radiation (TR), Cherenkov radiation (CR) etc. Along with transverse waves of radiation, 
the charged particle may also generate longitudinal oscillations. We show that all these phenomena may be described via quite simple and universal approach, 
where the source of the field is the polarization current density induced inside the medium by external field of the particle, that is direct proof of the physical equivalence of all these radiation processes. 
Exact solution for one of the basic radiation problems is found with this method: emission of a particle passing through a cylindrical channel in a screen of arbitrary width and permittivity
$\varepsilon (\omega ) = \varepsilon^{\prime} + i \varepsilon^{\prime \prime}$. Depending on geometry, the formula for radiated energy obtained describes different types
of polarization radiation: DR, TR and CR. The particular case of radiation produced by the particle crossing axially the sharp boundary between vacuum and a plasma cylinder of finite radius 
is also considered. The problem of SPR generated when the particle moves nearby a set of thin rectangular strips (grating) is solved for the arbitrary value of the grating's permittivity. 
An exact solution of Maxwell's equations for the fields of polarization current density suitable at the arbitrary distances (including the so-called pre-wave zone) is presented.
This solution is shown to describe transverse fields of polarization radiation and the longitudinal fields connected with the zeros of permittivity.
\end{abstract}




\maketitle

\section{\label{1}Introduction}

A charged particle moving nearby a spatially inhomogeneous condensed medium (target) or inside it may produce different types of radiation: 
Diffraction radiation, Smith-Purcell radiation, Cherenkov radiation, Transition radiation, Parametric X-ray radiation (PXR) etc. 
Along with transverse waves of radiation, the particle may also generate longitudinal waves. From the macroscopic point of view all these phenomena arise due to dynamic polarization of a medium by the external field of the moving charge and may be considered as the manifestation of the phenomenon known in the microscopic theory as \textit{polarization radiation} (PR) \cite{A}. 
Small radiation losses due to PR determine the wide range of its applications: from the new techniques of non-invasive beam diagnostics \cite{Kar, M, B, Doucas} 
to the new sources of THz radiation \cite{N, Kum, And, Leemans, Schroeder-2}.

However, up to now there is no universal theoretical description of all these radiation processes. The classical Ginzburg-Frank solution for TR is known to include CR contribution \cite{G-T}, 
but the corresponding solutions for DR and SPR generated on the surfaces of a finite permittivity (and therefore including possible CR) are still absent, even for the simplest targets.
Available results are bounded by TR and DR in X-ray range \cite{D, T}, where the value of permittivity is close to unit: $\varepsilon (\omega) - 1 \ll 1$, 
and the difference between various types of PR practically vanishes (for example, between CR and PXR \cite{Nas}). 
Moreover, in a real experiment one has different types of PR generated \textit{simultaneously}. The absence of possibility to estimate, for example, a contribution of DR to CR 
generated on the targets of finite dimensions hampers the use of them in beam diagnostics \cite{Tak, M}. The problem of coherent TR produced by the charged particles 
at the boundary of plasma occupying the finite space is of importance for development of new types of THz radiation sources \cite{Leemans, Schroeder-2}. But most available solutions are based on the presentation of the radiation from a sharp boundary between vacuum and plasma as from the ``vacuum - ideal conductor'' boundary (see, for example, \cite{Schroeder, Bosch-Plasma}).
Such a model is rather approximate, and a more rigorous consideration is required. Finally, some recent publications were devoted to demonstration of an analogy between the theory of TR and DR and the one of the wake fields in accelerators \cite{B, Stupakov, Xiang, X}. But there is still no good theory describing all the types of polarization radiation and the wake fields generated in a target of a given shape and finite permittivity $\varepsilon (\omega ) = \varepsilon^{\prime} + i \varepsilon^{\prime \prime}$.

The exact solution of Maxwell's equations describing all the types of radiation mentioned as polarization radiation, is necessary not only for the numerous applications of PR in accelerators, 
beam and plasma physics, but for more clear understanding of the general physical properties of PR itself. For instance, TR is usually studied solving \textit{homogeneous} Maxwell's eqs. with appropriate boundary conditions \cite{G-T, L}. Energy losses on Cherenkov radiation in an unbounded medium is obtained as a part of the total polarization losses corresponding to the excitation of transverse oscillations \cite{L, Ryaz}. In most cases, CR and TR are considered separately from each other (though there exist several works directly devoted to the investigation of interference between them), but several authors suppose that the mechanism of their generation is the same (unfortunately, without an adequate proof of this suggestion, see e.g. \cite{H}).
On the other hand, DR and SPR generated on the ideally conducting surfaces are often represented as the field of the surface current induced by external field of a particle 
(by analogy with diffraction of the ordinary plane waves, see e.g. \cite{Kaz, Til, Br, Kes}). Though this formulation seems to be correct, it was recently shown that diffraction of the particle's field 
(not the one of a plane-wave) cannot be described with the use of a surface current with only two tangential components \cite{PLA}. 
It turned out, that exact expression for the surface current density induced by the particle's field on an ideally conducting surface has all three components including the one perpendicular to the screen (generalized surface current \cite{PLA}). 

Due to the fact that such different ways are used to obtain the formulas for different types of the same polarization radiation, \textit {the physical origin} of these types of radiation is often just not indicated.
In general, it is not enough to say that CR arises when the particle velocity is larger than that of light in this medium, as many authors still do. It is necessary to indicate \textit {the source of radiation}, 
that is polarization current density in this case (see below). The well-known Cherenkov relation between angle of emission, particle velocity and permittivity is just a condition of the constructive interference of waves emitted at each point of the particle trajectory. The problem of the physical equivalence (or non-equivalence) of TR and CR eventually is connected with the fact, that exactly the source of radiation is often not indicated (because these are homogeneous equations that are solved in the case of TR). In what follows, we shall present the source of radiation leading directly to the well-known formulas for TR intensity including possible CR contribution, that is direct proof of their physical equivalence. The term ``equivalence'' here should be understood in the way, that both TR and CR, as well as DR, SPR and the longitudinal fields, physically arise due to polarization of a medium by the field of the moving particle. Therefore, the difference between these radiation processes is completely ``kinematic'', and just the presence of interface of some shape leads to the changes in the radiation characteristics.

The organization of the paper is as follows: in Sec.\ref{Sect2} we find the exact solution of Maxwell's equation for magnetic field of polarization current density induced inside a target of a given shape
and permittivity by the field of external source and the field of PR itself. In Sec.\ref{Sect3} we consider the problem of radiation generated by the particle moving through cylindrical channel inside the layer of finite width and permittivity. It turns out that the energy radiated through the plane interface into vacuum may be easily found using the well-known reciprocity theorem \cite{L}. 
In Sec.\ref{Sect3.5} we find the solution for the problem of polarization radiation produced by a particle crossing the boundary between vacuum and a plasma cylinder of finite radius.
Sec.\ref{Sect4} is devoted to solution of the Smith-Purcell radiation problem arising while the particle moves nearby a thin grating of finite permittivity. 
Finally, in Appendix the exact expressions for the fields of polarization current density applicable at the arbitrary distances (including the so-called pre-wave zone) are derived and analyzed.

\section{\label{Sect2} Polarization radiation field inside a medium}

When a charged particle moves nearby a condensed medium or inside it, its own field polarizes atoms and molecules, and the secondary field known as polarization radiation arises. 
In the microscopic theory this phenomenon is often called ``atomic bremsstrahlung'' (see e.g. \cite{A} and references therein). This secondary field propagating inside the medium in the form 
of transverse wave or exciton also polarizes atoms, that is why the source of PR is the polarization current density linearly depending upon \textit{the total field} including external field of the moving particle $\bold E^0$ and the field $\bold E^{pol}$ of PR itself (the medium is non-magnetic throughout the paper: $\mu = 1$; Gaussian system of units is used):
\begin{equation}
\displaystyle \bold {j}_{pol} = \sigma (\bold E^0 + \bold E^{pol} (\bold {j}_{pol})). \label{Eq1}
\end{equation}
Note that this is the conventional expression for the macroscopic current density induced inside a medium used when averaging microscopic Maxwell's eqs. (see e.g. \cite{L, Ryaz}).
Here conductivity is a given function of permittivity possessing frequency dispersion: 
\begin{eqnarray}
\displaystyle && \sigma (\bold r, \omega) = \frac{\omega}{4 \pi i}(\varepsilon (\bold r, \omega) - 1) . \label{Eq8.5}
\end{eqnarray}  
For non-magnetic media, Eq.(\ref{Eq1}) just means that the field of polarization radiation is generated by electric dipole moments with effective volume density $\bold P (\bold r, \omega) =
\bold j_{pol} (\bold r, \omega)/(- i \omega) = (\varepsilon - 1) (\bold E^0 + \bold E^{pol})/(4 \pi)$.

Equation (\ref{Eq1}) is obviously integral, and neglect of the integral term $\bold E^{pol} (\bold {j}_{pol})$ is possible only for X-ray radiation, when the secondary re-emissions may be ignored
due to smallness of parameter $\varepsilon (\omega) - 1 \ll 1$ \cite{D, T}. However we can get rid of this term replacing it to the left side of ``vacuum'' Maxwell's equations, that leads to the replacement $\omega/c \rightarrow \sqrt{\varepsilon (\bold r, \omega)} \omega/c$. Maxwell's equations in this case may be reduced to the form:
\begin{eqnarray}
\displaystyle \Big (\Delta + \varepsilon (\bold r, \omega) \frac{\omega^2}{c^2}\Big ) \bold H^{pol} (\bold r, \omega ) = - \frac{4 \pi}{c}
\Big (\sigma (\bold r, \omega) \curl \bold E^0 \cr \qquad - (\bold E^0 + \bold E^{pol}) \times \grad \sigma (\bold r, \omega) \Big ), \label{Eq2}
\end{eqnarray}
which still contains $\bold E^{pol}$ in the right side.
\begin{figure}
\center \includegraphics[width=5.50cm, height=5.00cm]{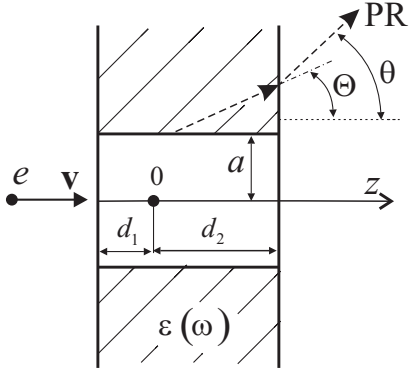}
\caption{\label{Fig1} Generation of PR by a charged particle moving through a channel in the layer}
\end{figure}

If the spatial dependence of conductivity $\sigma (\bold r, \omega)$ is step-like: $\sigma (\bold r,\omega) = \Theta (z) \sigma (\omega)$ (that corresponds to the plane interface between vacuum and medium) or has the more complicated form corresponding to a target of a given shape, it is easy to see that the last term in the right side of (\ref{Eq2}) may be expressed 
through the boundary conditions on the interface. For example, in the former case ($\bold n$ is a unit normal to the interface):
\begin{eqnarray}
&& \displaystyle (\bold E^0 + \bold E^{pol})\times \grad \sigma (\bold r, \omega) = \sigma (\omega) \delta(z) \bold E^0 \times \bold n, \label{Eq3}
\end{eqnarray}
i.e. tangential components of electric field are continuous on an interface. Therefore solution of Eq.(\ref{Eq2}) may be written straightforward:
\begin{eqnarray}
\displaystyle \bold H^{pol} (\bold r, \omega ) = \curl \frac{1}{c} \int \limits_{V_T} \bold \sigma (\omega) \bold E^0 (\bold {r}^{\prime}, \omega) 
\frac{e^{i \sqrt{\varepsilon (\omega )} \omega |\bold r - {\bold r}^{\prime}|/c}}{|\bold r - {\bold r}^{\prime}|} d^3 r^{\prime}, \label{Eq4}
\end{eqnarray}
where the region of integration is reduced only to the volume of the medium (target) $V_T$. An exact expression for electric field strength and its analysis is provided in Appendix.

Note that this expression is the \textit{exact solution} of Maxwell's eqs. with current density (\ref{Eq1}) in the right side. The integral term in the right-hand side of (\ref{Eq1}) resulted only in appearance of $\sqrt{\varepsilon (\omega)}$ in exponent. Physically, such procedure means ``renormalization'' of the field acting on each atom and molecule because of the presence of other atoms 
(that is macroscopically characterized by the permittivity $\varepsilon (\omega)$) and leads to the replacement 
\begin{eqnarray}
\displaystyle
\bold E^{total} \rightarrow \bold E^{0}, \ \omega/c \rightarrow \sqrt{\varepsilon (\omega)} \omega/c, \label{Eq4h}
\end{eqnarray}
as formula (\ref{Eq4}) shows. The last is correct for any shape of the target, and in spite of its rather simple form describes \textit{all the types} of PR inside a medium (see below).

To demonstrate how this approach works, we consider one of the basic radiation problems: emission of a charged particle moving uniformly along the axis of the round cylindrical channel 
with radius $a$ in a screen of finite thickness $d = d_1 + d_2$ and arbitrary permittivity $\varepsilon (\omega ) = \varepsilon^{\prime} + i \varepsilon^{\prime \prime}$ (see Fig.\ref{Fig1}). 
In this geometry two types of PR arise: DR and CR, but if the hole radius approaches zero DR ought to transform into TR. Let us check these considerations. 
According to (\ref{Eq4}) PR field in the wave zone is found as (vector $\bold H^{R}$ denotes the far-field part of $\bold H^{pol}$, see Appendix):
\begin{widetext}
\begin{eqnarray}
\displaystyle && \bold H^{R} (\bold r, \omega ) = \frac {e \omega^2 (\varepsilon - 1)}{4 \pi^2 v^2 c \gamma} \frac{e^{i r \sqrt{\varepsilon} \omega/c}}{r} \bold k \times 
\int \limits_{a}^{\infty}\rho^{\prime}d\rho^{\prime} \int \limits_{0}^{2 \pi} d\phi^{\prime} \int \limits_{-d_1}^{d_2} dz^{\prime}
\Big(\frac{{\bm \rho^{\prime}}}{\rho^{\prime}} K_1 \Big[\frac {\omega \rho^{\prime}}{v \gamma}\Big] - \displaystyle \frac {i}{\gamma}\frac {\bold v}{v}
K_0 \Big[\frac{\omega \rho^{\prime}}{v \gamma}\Big]\Big) e^{- i \bold k \bold r^{\prime} + i \frac {\omega}{v} z^{\prime}}. \label{Eq5}
\end{eqnarray}
\end{widetext}
Here we use conventional expression for the field of uniformly moving charge $e$ with energy $\gamma = 1/\sqrt{1 - \beta^2}$ \cite{Jakson} (our coefficient in definition of Fourier transform differs from that in \cite{Jakson}, but the final result for radiated energy does not depend on it of course). In (\ref{Eq5}) also: $K_{0,1}$ are McDonald functions, $\bold {r}^{\prime} = \{\bm \rho^{\prime}, z^{\prime}\} = \{\rho^{\prime} \cos {\phi^{\prime}}, \rho^{\prime} \sin {\phi^{\prime}}, z^{\prime}\}$ is the vector of volume integration,
$\bold v = \{0, 0, v\}$ is the particle velocity vector, and the wave vector in medium is $\bold {k} = \frac{\bold r}{r}\sqrt{\varepsilon(\omega)}\omega/c$, where
\begin{eqnarray}
\displaystyle && \bold {r} = r \{\sin{\Theta} \cos {\phi}, \sin{\Theta} \sin {\phi}, \cos{\Theta}\}. \label{Eq5.1}
\end{eqnarray}

Integration in (\ref{Eq5}) is performed using well-known presentations for cylindrical functions
\cite{R} (see also such integrals in Ref.\cite{J}), and the final result for PR field is:
\begin{widetext}
\begin{eqnarray}
\displaystyle && \bold H^{R} (\bold r, \omega ) = \frac {e \omega \sqrt{\varepsilon} (\varepsilon - 1) a}{2 \pi v c \gamma^2} \frac{e^{i r \sqrt{\varepsilon} \omega/c}}{r} 
\frac{e^{-i d_1 \frac{\omega}{c}(\beta^{-1} - \sqrt{\varepsilon} \cos{\Theta})}- e^{i d_2 \frac{\omega}{c} (\beta^{-1} - \sqrt{\varepsilon} \cos{\Theta})}}{(\beta^{-1} - \sqrt{\varepsilon} \cos{\Theta})(1 - \beta^2 + (\beta \sqrt{\varepsilon} \sin{\Theta})^2)}\{\sin {\phi}, - \cos {\phi}, 0\} \cr \displaystyle && \ \times \Big ( \sin{\Theta} (\gamma^{-1} - \beta \gamma \sqrt{\varepsilon}
\cos{\Theta}) J_0\Big (a \frac{\omega}{c} \sqrt{\varepsilon} \sin{\Theta}\Big ) 
K_1\Big (a \frac{\omega}{v \gamma}\Big ) - (\cos{\Theta} + \beta \sqrt{\varepsilon} \sin^2{\Theta}) J_1\Big (a \frac{\omega}{c} \sqrt{\varepsilon} \sin{\Theta}\Big ) 
K_0\Big (a \frac{\omega}{v \gamma}\Big )  \Big ), \label{Eq6}
\end{eqnarray}
\end{widetext}
where $J_{0,1}$ are the Bessel functions.

First of all, we check the case of homogeneous medium: $d_1, d_2\rightarrow \infty$. Interference term is transformed as:
\begin{eqnarray}
\displaystyle e^{-i d_1 \frac{\omega}{c}f}- e^{i d_2 \frac{\omega}{c}f}\rightarrow i f \int \limits_{-\infty}^{\infty}e^{i f z} dz= 2 \pi i f \delta (f). \label{Eq7}
\end{eqnarray}
As $f = \beta^{-1} - \sqrt{\varepsilon} \cos{\Theta}$, the zero of delta-function gives the ordinary \textit{Cherenkov condition}:
\begin{eqnarray}
\displaystyle && \cos{\Theta} = \frac{1}{\beta \sqrt{\varepsilon}}, \label{Eq5.2}
\end{eqnarray}
and the formula (\ref{Eq6}) describes CR generated in a channel in the unbounded medium \cite{Bol}. 

In the limit $a\rightarrow 0$ the following relations are fulfilled (arguments of Bessel functions are omitted) \cite{R}:
\begin{eqnarray}
\displaystyle && a J_0 K_1 \rightarrow \frac{v \gamma}{\omega}, \ a J_1 K_0 \rightarrow 0, \label{Eq10}
\end{eqnarray}
and the energy radiated into unit frequency interval is found using (\ref{Eq6}) as (the medium is transparent):
\begin{widetext}
\begin{eqnarray}
\displaystyle \frac{d W}{d \omega} = \int  \frac{c r^2}{\sqrt{\varepsilon}}|\bold H^{R}|^2 d \Omega = \frac{2 e^2 \beta^2}{\pi c}\sqrt{\varepsilon} 
\int \limits_0^\pi \frac{\sin^2{\Big (\frac{d}{2} \frac{\omega}{v} (1 - \beta \sqrt{\varepsilon} \cos{\Theta})\Big )}}{(1 - \beta \sqrt{\varepsilon} \cos{\Theta})^2} 
\Big [\frac{(\varepsilon - 1) (1 - \beta^2 - \beta \sqrt{\varepsilon} \cos{\Theta})}{1 - \beta^2 + (\beta \sqrt{\varepsilon} \sin{\Theta})^2} \Big ]^2 \sin^3 {\Theta} d \Theta. \label{Eq6b}
\end{eqnarray}
\end{widetext}
If the Cherenkov condition is fulfilled, the term in the square brackets is equal to unit, and the formula (\ref{Eq6b}) is reduced to the one known in the Tamm theory (see e.g. \cite{Z}).
For the layer of large width, interference term transforms into delta-function:
\begin{eqnarray}
\displaystyle && \frac{\sin^2{\Big (\frac{d}{2} \frac{\omega}{v} (1 - \beta \sqrt{\varepsilon} \cos{\Theta})\Big )}}{(1 - \beta \sqrt{\varepsilon} \cos{\Theta})^2} \rightarrow \cr && \qquad \qquad \rightarrow
\frac{d}{2}\frac{\omega}{v} \pi \ \delta \Big (1 - \beta \sqrt{\varepsilon} \cos{\Theta}\Big ). \label{Eq6c}
\end{eqnarray}
After this transformation integration in (\ref{Eq6b}) gives
\begin{eqnarray}
\displaystyle && \frac{1}{d}\frac{d W}{d \omega} = \frac{e^2}{c^2}\omega \Big (1 - \frac{1}{\beta^2 \varepsilon(\omega)}\Big ) \label{Eq6d}
\end{eqnarray}
that is the ordinary formula for CR spectral density in the unbounded medium (see e.g. \cite{L}).

\section{\label{Sect3}The energy emitted into vacuum}

Formula (\ref{Eq6}) has been derived for polarization radiation field taken inside a medium. In order to find radiation intensity outside the target (in vacuum), 
we cannot use the ordinary Fresnel laws of refraction because for absorbing medium the radiating dipoles are concentrated in the vicinity of interface and the radiation field in this region 
does not correspond to the far-field (it is always so if $\varepsilon^{\prime \prime}\sim \varepsilon^{\prime}$). To solve this problem, it is useful to apply the so-called reciprocity theorem \cite{L}.
The idea is in solving the \textit{inverse problem}, we are to find the field refracted into medium if the field of the wave incident from vacuum onto $xOy$ plane 
is determined by Eq.(\ref{Eq6}). In this case it is possible to use the ordinary Fresnel laws of refraction, and finding the solution for such inverse problem we can solve the initial problem, i.e. obtain the field refracted from medium into vacuum. Thanks to azimuthal symmetry, we find the simple relation (in more detail see Ref.\cite{JETPL}):
\begin{eqnarray}
\displaystyle && |\bold E^R_{vac}|^2 = \Big | \frac{f_H}{\varepsilon}\Big |^2 |\bold H^R|^2 = \cr && \qquad \qquad = \Bigg |\frac{2 \cos {\theta}}{\varepsilon \cos {\theta} + \sqrt{\varepsilon - \sin^2{\theta}}}\Bigg |^2 |\bold H^R|^2, \label{Eq8}
\end{eqnarray}
where 
\begin{eqnarray}
\displaystyle && \sin {\theta} = \sqrt{\varepsilon}\sin{\Theta}, \label{Eq10.1}
\end{eqnarray}
$\theta$ is the polar angle in vacuum \cite{L}, and $f_H$ is the Fresnel coefficient. Note that no additional approximations were made when deriving Eq.(\ref{Eq8}). 
It should be also emphasized that though the use of a polar angle in medium $\Theta$ has a sense only for transparent media, all the results below have no restrictions on the imaginary part of permittivity.

Thus in order to find the energy radiated in a unit solid angle and frequency interval according to relation $d^2W/d\omega d\Omega = c r^2 |\bold E^R_{vac}|^2$, it is enough to express formula (\ref{Eq6}) through the vacuum angle $\theta$. 
Doing this and supposing $d_1 = d, d_2 = 0$ we get finally:
\begin{widetext}
\begin{eqnarray}
\displaystyle \frac{d^2W}{d\omega d\Omega} = \frac{e^2}{\pi^2 c}  \frac{\beta^2 \cos^2{\theta}}{(1 - \beta^2 \cos^2{\theta})^2} 
\Big | e^{-i d \frac{\omega}{c}\Big (\beta^{-1} - \sqrt{\varepsilon - \sin^2{\theta}}\Big )}- 1\Big |^2 
\Big (a \frac {\omega}{v\gamma}\Big )^2 \Bigg |\frac{\varepsilon - 1}{(1 - \beta \sqrt{\varepsilon - \sin^2{\theta}})(\varepsilon \cos {\theta} + \sqrt{\varepsilon - \sin^2{\theta}})}
 \times\cr \displaystyle \Big (\sin{\theta} (1 - \beta^2 - \beta \sqrt{\varepsilon - \sin^2{\theta}}) J_0\Big (a \frac{\omega}{c} \sin{\theta}\Big ) K_1\Big (a \frac{\omega}{v \gamma}\Big )  - \gamma^{-1}(\sqrt{\varepsilon - \sin^2{\theta}} + \beta \sin^2{\theta}) J_1\Big (a \frac{\omega}{c} \sin{\theta}\Big ) K_0\Big (a \frac{\omega}{v \gamma}\Big ) \Big )\Bigg |^2 \label{Eq9}
\end{eqnarray}
\end{widetext}
The expression derived describes \textit{all the types} of PR generated in a layer and emitted into vacuum: CR, DR or TR (if $a=0$). 

In the limiting case of semi-infinite continuous medium ($a \rightarrow 0$ and $d \rightarrow \infty$) the formula (\ref{Eq9}) is reduced to the well-known Ginzburg-Frank expression 
for forward TR \cite{G-T, L}:
\begin{widetext}
\begin{eqnarray}
&& \displaystyle \frac{d^2W}{d\omega d\Omega} \rightarrow \frac{e^2}{\pi^2 c}  \frac{\beta^2 \sin^2 {\theta}\cos^2{\theta}}{(1 - \beta^2 \cos^2{\theta})^2} 
\Bigg |\frac{(\varepsilon - 1)(1 - \beta^2 - \beta \sqrt{\varepsilon - \sin^2{\theta}})}
{(1 - \beta \sqrt{\varepsilon - \sin^2{\theta}})(\varepsilon \cos {\theta} + \sqrt{\varepsilon - \sin^2{\theta}})}\Bigg |^2. \label{Eq11}
\end{eqnarray}
\end{widetext}
Here (as well as in (\ref{Eq9})) the pole $\beta \sqrt{\varepsilon - \sin^2{\theta}} = 1$ corresponds to CR cone refracted into vacuum,
but for the target of finite width this pole is removable (the medium is transparent):
\begin{eqnarray}
&& \displaystyle \Big | \frac{e^{-i d \frac{\omega}{c}(\beta^{-1} - \sqrt{\varepsilon - \sin^2{\theta}})}- 1}{1 - \beta \sqrt{\varepsilon - \sin^2{\theta}}}\Big |^2 = 
\cr && \ = \frac{4 \sin^2{\Big (\frac{d}{2} \frac{\omega}{c} (\beta^{-1} - \sqrt{\varepsilon - \sin^2{\theta}})\Big )}}{(1 - \beta \sqrt{\varepsilon - \sin^2{\theta}})^2} \rightarrow \Big (\frac{\omega}{v}d\Big )^2, \label{Eq12}
\end{eqnarray}
and the radiation intensity is proportional to the width squared, as in Tamm theory of CR without channel (see e.g. \cite{Bol, Z} and references therein). 
Note that integration over angles leads to the linear dependence on $d$.

The argument of $\sin^2(d/d_f)$ in (\ref{Eq12}) is related to DR (TR) \textit{formation length} in the weakly absorbing medium: 
\begin{eqnarray}
\displaystyle && d_f = \frac{\lambda}{\pi |\beta^{-1} - \sqrt{\varepsilon - \sin^2{\theta}}|}, \label{Eq8.3}
\end{eqnarray} 
becoming infinity for the ``pure'' CR (when $\varepsilon^{\prime \prime} = 0$). If the target thickness $d$ is much less than $d_f$, we get for TR the same $\propto d^2$ dependence even if Cherenkov condition is not fulfilled. If the opposite condition holds true $d \gg d_f$ (thick target), the interference term again transforms into delta-function according to (\ref{Eq6c}), determining the Cherenkov condition via the vacuum angle $\theta$ (because in this case CR is observed in vacuum, but not in the medium as in the previous paragraph). 
Far from the Cherenkov angle, $\sin^2(d/d_f)$ for the thick target is highly oscillating and may be replaced by its mean value $1/2$. 
So, in contrast to CR the intensity of transition radiation (diffraction radiation) in a transparent medium almost does not depend on the target thickness when condition $d \gg d_f$ is fulfilled, see Fig. 2.
\begin{figure}
\center \includegraphics[width=9.00cm, height=5.50cm]{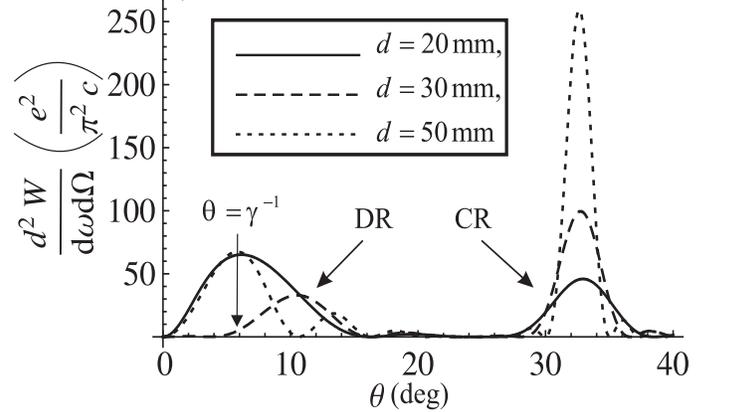}
\caption{\label{Fig2}Angular distributions of the total polarization radiation in vacuum. Parameters: $\gamma = 10, \lambda = 1$mm, $a = 1$mm, $\varepsilon^{\prime} = 1.3$,
$\varepsilon^{\prime \prime} = 10^{-3}$.}
\end{figure}
\begin{figure*}
\center \includegraphics[width=15.00cm, height=5.20cm]{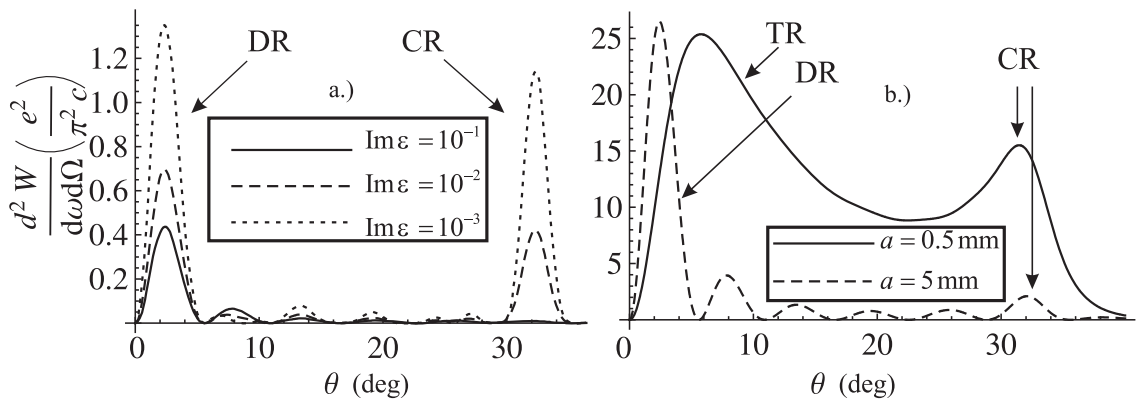}
\caption{\label{Fig3}Angular distributions of the total polarization radiation in vacuum. Parameters: $\gamma = 10, \lambda = 1$mm, $d = 40$mm; a.) $a = 5$mm, $\varepsilon^{\prime} = 1.3$;
b.) $\varepsilon = 1.3 + i 0.05$ (dashed curve is multiplied by the factor 60). The angle of Cherenkov radiation reflected into vacuum is determined from the condition
$\beta \sqrt{\varepsilon - \sin^2{\theta}} = 1$.}
\end{figure*}

In another limiting case of conducting target with $\varepsilon^{\prime \prime} \gg \varepsilon^{\prime}$, exponential term in (\ref{Eq9}), (\ref{Eq12}) becomes damped, and dependence upon the target width disappears if condition $d \gg \lambda/\sqrt{|\varepsilon|}$ is fulfilled (\textit{skin-effect}). In the case of ideal conductivity ($\varepsilon^{\prime \prime}\rightarrow \infty$) formula (\ref{Eq9}) describes only forward DR:
\begin{widetext}
\begin{eqnarray}
\displaystyle \frac{d^2 W}{d\omega d\Omega} \rightarrow \frac{e^2}{\pi^2 c}
\frac{\beta^2 \sin^2{\theta}}{(1 - \beta^2 \cos^2{\theta})^2} \Big (\frac{a \omega}{v \gamma} \Big )^2 \Big (J_0 \Big (a \frac{\omega}{c} \sin{\theta}\Big ) 
K_1 \Big (a \frac{\omega}{v \gamma} \Big ) + \frac{1}{\beta \gamma \sin{\theta}} J_1 \Big (a \frac{\omega}{c} \sin{\theta}\Big ) K_0 \Big (a \frac{\omega}{v \gamma} \Big ) \Big )^2, \label{Eq13}
\end{eqnarray}
\end{widetext}
This is exact expression for DR generated by a particle moving through a round aperture in an ideally-conducting screen, that was found with another methods e.g. in Refs. \cite{J, X}.

Note that formula (\ref{Eq9}) describes only forward polarization radiation. The corresponding expression for \textit{backward} emission is found 
by the substitution $\cos {\Theta}\rightarrow - \cos {\Theta}$ in (\ref{Eq6}) and applying the same refraction rule (\ref{Eq8}), see also \cite{JETPL}.

In the general case, when the particle moves through the channel in the screen with arbitrary permittivity $\varepsilon (\omega) = \varepsilon^{\prime} + i \varepsilon^{\prime \prime}$, 
formula (\ref{Eq9}) completely describes this process, i.e. represents the \textit{total} radiation intensity including Cherenkov and diffraction mechanisms.
It should be noted that we consider refraction of polarization radiation from medium into vacuum through the flat interface laying in $xOy$ plane (see Fig.1), so we do not take into account the secondary reflections of the emitted radiation inside the channel, that seems to be relevant when the screen width $d$ is large enough. However this approximation is in fact insignificant, 
because Cherenkov radiation as is well known is not radiated through the interface parallel to the particle trajectory (see below). In other words, the total internal reflection takes place, 
and in this case CR is emitted into vacuum only through the surface perpendicular to the particle trajectory (if CR condition $\sin^2 \theta = |\varepsilon - 1/\beta^{2}| < 1$ is fulfilled), see e.g. \cite{Bol}.
\begin{figure}
\center \includegraphics[width=8.50cm, height=5.00cm]{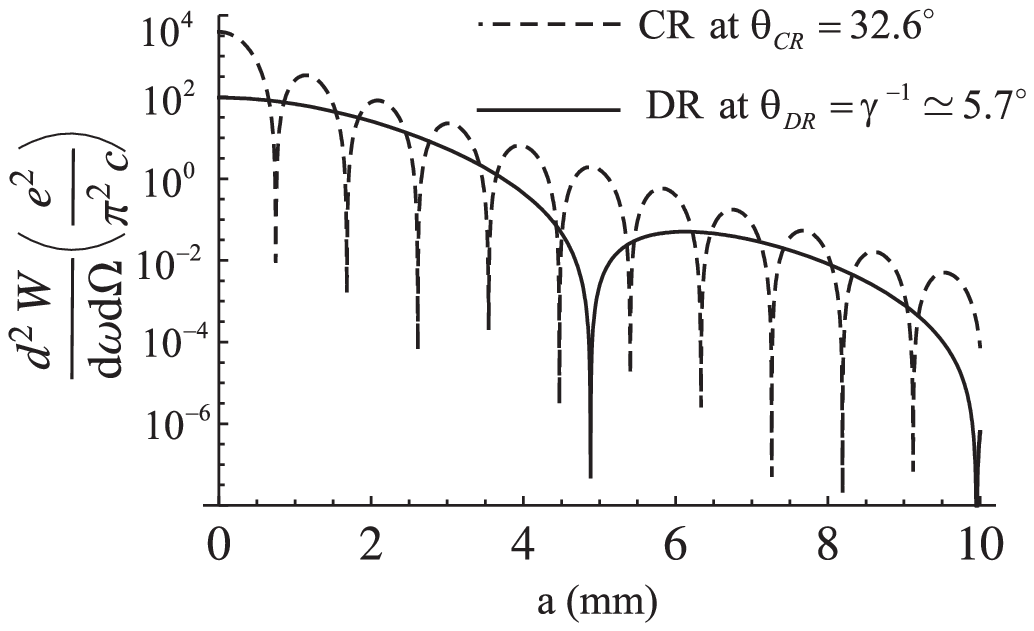}
\caption{\label{Fig4}Dependence of polarization radiation intensity on the hole radius. Parameters: $\gamma = 10, \lambda = 1$mm, $d = 50$mm, $\varepsilon^{\prime} = 1.3$, $\varepsilon^{\prime \prime} = 10^{-3}$.}
\end{figure}
\begin{figure}
\center \includegraphics[width=8.50cm, height=5.00cm]{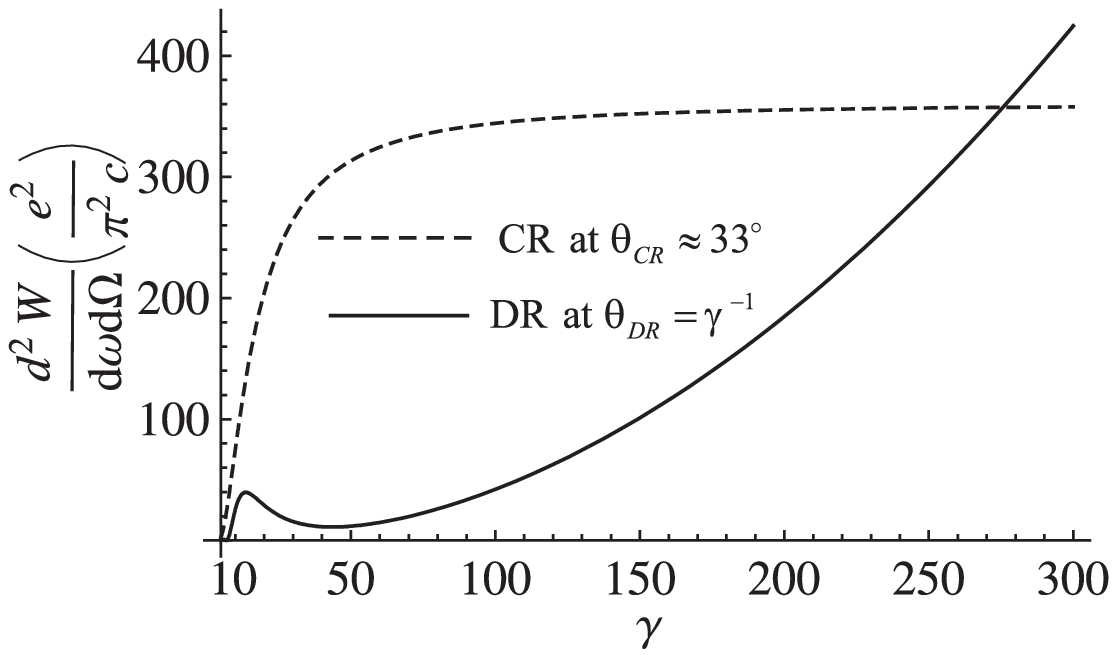}
\caption{\label{Fig5}Dependence of polarization radiation intensity on the particle energy. Parameters: $\lambda = 1$mm, $a = 2$mm, $d = 50$mm, $\varepsilon^{\prime} = 1.3$, $\varepsilon^{\prime \prime} = 10^{-3}$.}
\end{figure}

The angular and other distributions of the energy losses due to different types of polarization radiation are shown in Figs.\ref{Fig3},\ref{Fig4},\ref{Fig5}. 
As may be seen in Fig.\ref{Fig3}a.), increase of imaginary part of permittivity leads to CR suppression. Increase of the channel radius results in transform of TR into DR and in decrease of CR contribution, see Fig.\ref{Fig3}b.). Fig\ref{Fig4} shows, that the presence of $\sin \theta$ in the argument of the oscillating Bessel functions in Eq.(\ref{Eq9}) leads to the different dependence on $a$ 
for intensity of diffraction radiation (small $\theta$) and the one of Cherenkov radiation (large $\theta$). The difference between them reveals itself 
also in the dependence of the radiated energy on the Lorentz factor of the particle, see Fig.\ref{Fig5}. In particular, this figure shows that for the large enough values of $\gamma$ 
(actually, when $\gamma \lambda/(2 \pi) \gg a$), energy dependence of DR becomes quadratic as for TR \cite{G-T}.    

It is clear from the figures that all three types of radiation (TR, DR and CR) have the same physical origin, and the boundary between them is rather relative. 
Note that the physical equivalence of TR and DR was shown for an ideally conducting target in the ultrarelativistic limit relatively recently \cite{Pap}. 
Here we proved this for the case of arbitrary permittivity and arbitrary particle energy that allowed to consider possible contribution of Cherenkov radiation.

\section{\label{Sect3.5} Polarization radiation from a plasma cylinder}

Study of radiation generated by the charged particles crossing the boundary between a channel of the dense plasma and vacuum is of importance for development of new THz radiation sources
\cite{Leemans, Schroeder-2}. TR from such a boundary was modeled in Refs.\cite{Schroeder, Bosch-Plasma} as the one produced at the boundary between vacuum and an ideal conductor.
This was justified by the fact that the value of the electron plasma permittivity
\begin{eqnarray}
&& \displaystyle \varepsilon (\omega ) \approx 1 - \frac{\omega_{p(e)}^2}{\omega^2}, \label{Eq3.5.1}
\end{eqnarray}
tends to $-\infty$ when $\omega_{p(e)} \gg \omega$ (dense plasma, see Ref.\cite{Schroeder}). 

Actually, for frequency range below the optical band, this formula is not applicable for collisionless plasma, because the effects of the spatial dispersion of permittivity are ignored.
But even neglecting by the spatial dispersion, this formula works only for small densities ($\omega_{p(e)}$ is of order or less than $\omega$), because it is based on the supposition of the linear dependence of permittivity on the density of electrons $n_e$:
\begin{eqnarray}
&& \displaystyle \varepsilon (\omega ) = 1 + 4 \pi n_e \alpha_e (\omega ), \label{Eq3.5.2}
\end{eqnarray}
where $\alpha_e (\omega )$ is the polarizability of electrons (electrons in plasma or conduction electrons in a metal, see e.g. problem 11.17 in \cite{Toptygin}). 
In the simplest model of oscillators with zero own frequency (cold plasma) and attenuation factor $\Gamma_e$, the polarizability of such electrons is (see e.g. \cite{Ryaz, Toptygin})
\begin{eqnarray}
&& \displaystyle \alpha_e (\omega ) = -\frac{e^2}{m_e} \frac{1}{\omega^2 + i \Gamma_e \omega}, \label{Eq3.5.3}
\end{eqnarray}
If $\Gamma_e \rightarrow 0$ (i.e. neglecting by Landau damping), and with notation $\omega_{p(e)}^2 = 4 \pi n_e e^2/m_e$, the formula (\ref{Eq3.5.2}) coincides with (\ref{Eq3.5.1}).

The more rigorous study allowing to consider high densities is based on Clausius-Mossotti formula for an electrically neutral isotropic medium of weakly interacting particles:
\begin{eqnarray}
&& \displaystyle \varepsilon (\omega ) = \frac{1 + \frac{8 \pi}{3} n \alpha (\omega )}{1 - \frac{4 \pi}{3} n \alpha (\omega )} = 1 + \frac{4 \pi n \alpha (\omega)}{1 - 
\frac{4 \pi}{3} n \alpha (\omega )}. \label{Eq3.5.4}
\end{eqnarray}
Here $n \alpha (\omega ) = \sum \limits_s n_s \alpha_s (\omega )$ represents the sum over all subsystems of the medium being studied. 
For electrically neutral (as a whole) plasma with polarizability of electrons (\ref{Eq3.5.3}), zero damping and heavy ions: $m_i \gg m_e$, the last relation is reduced to (\ref{Eq3.5.2}) 
under the condition of small density
\begin{eqnarray}
&& \displaystyle n \ll \frac{3}{4 \pi r_e} \frac{\omega^2}{c^2}, \label{Eq3.5.5}
\end{eqnarray}
or high frequency
\begin{eqnarray}
&& \displaystyle \omega \gg \omega_{p(e)}, \label{Eq3.5.5-2}
\end{eqnarray}
where $r_e = e^2 /(m_e c^2)$ is the classical electron radius. As the opposite case $\omega \ll \omega_{p(e)}$ does not led to (\ref{Eq3.5.2}), it is impossible to apply the formula (\ref{Eq3.5.1}) 
for high density plasma.

Consider the partially ionized electrically neutral medium consisting of free electrons, bound electrons and ions. Polarizability of free electrons is determined by Eq.(\ref{Eq3.5.3}). 
The analogous expression for bound electrons takes into account the possible own frequency of the oscillators $\omega_0$ (only one for simplicity):
\begin{eqnarray}
&& \displaystyle \alpha_{b.e.} (\omega ) = -\frac{e^2}{m_e} \frac{1}{\omega^2 + i \Gamma_i \omega - \omega_0^2}, \label{Eq3.5.6}
\end{eqnarray}
Since $\alpha_i \propto m_i^{-1}$, we shall neglect by the contribution of ions due to their big mass: $m_i \gg m_e$.
So the formula (\ref{Eq3.5.4}) yields:
\begin{eqnarray}
&& \displaystyle \varepsilon (\omega ) = \frac{1 - \displaystyle \frac{2}{3}\frac{\omega_{p(e)}^2}{\omega^2 + i \Gamma_e \omega} - \displaystyle \frac{2}{3}\frac{\omega_{p(i)}^2}
{\omega^2 + i \Gamma_i \omega - \omega_0^2}}{1 + \displaystyle \frac{1}{3}\frac{\omega_{p(e)}^2}{\omega^2 + i \Gamma_e \omega} + \displaystyle \frac{1}{3}\frac{\omega_{p(i)}^2}
{\omega^2 + i \Gamma_i \omega - \omega_0^2}}, \label{Eq3.5.7}
\end{eqnarray}
where $\omega_{p(i)}^2 = 4 \pi n_i e^2/m_e$ is denoted, and $n_i$ is the density of ions with bound electrons. If the term $\omega^2 + i \Gamma_i \omega - \omega_0^2$ 
does not turn into zero, and the density of the partially recombined ions $n_i$ is much less than the one of free electrons $n_e$, 
the expression for permittivity describes only electron plasma:
\begin{eqnarray}
&& \displaystyle \varepsilon (\omega ) \approx 1 - \frac{\omega_{p(e)}^2}{\omega^2 + \displaystyle \frac{1}{3}\omega_{p(e)}^2 + i \Gamma_e \omega}, \label{Eq3.5.8}
\end{eqnarray}
Here, the term $\frac{1}{3}\omega_{p(e)}^2$ in the denominator appears because of the taking into consideration the so-called \textit{local field} effect in Clausius-Mossotti formula.
In fact, for high density of a medium (i.e. when condition (\ref{Eq3.5.5}) is not fulfilled) the field acting on each atom or an electron in plasma does not coincide with the commonly used macroscopic field $\bold E = \bold E^0 + \bold E^{pol}$, and dependence of permittivity on the density becomes non-linear as Eq.(\ref{Eq3.5.4}) shows (see also \cite{Ryaz, Toptygin}). 
In this case, even the formal transition $\omega_{p(e)} \rightarrow \infty$ in (\ref{Eq3.5.8}) (having no physical sense whatsoever, because the formulas (\ref{Eq3.5.2}), (\ref{Eq3.5.4}) were derived for a set of weakly interacting particles) results in the limit value of permittivity $\varepsilon (\omega) \rightarrow -2$, but not $- \infty$. 
It means that the modeling of high density plasma as the ideal conductor used in Refs.\cite{Schroeder, Bosch-Plasma} appears as the rather rude approximation. 
On the other hand, for the special problem of PR generated by a relativistic particle in a plasma cylinder, the difference between the radiation characteristics
for the value of permittivity $\varepsilon \sim -1.5$ and for the case of ideal conductor ($\varepsilon = i \infty$) is significant only for large angles of emission $\theta \gg \gamma^{-1}$ (see below).

Consider an electron crossing axially the sharp boundary between the plasma cylinder of radius $a$ and vacuum. The forward transition radiation may undergo diffraction 
due to the finite radius of cylinder. Such a problem is ``complementary'' to the one has been solved in the previous Sections. 
So, the radiation field inside the cylinder is found using Eq.(\ref{Eq5}) (integrating over $\rho^{\prime}$ from $0$ to $a$) as:
\begin{widetext}
\begin{eqnarray}
\displaystyle && \bold H^{R} (\bold r, \omega ) = \bold H^{R}_{TR} - (\ref{Eq6}) = \frac {e \sqrt{\varepsilon} (\varepsilon - 1)}{2 \pi c} \frac{e^{i r \sqrt{\varepsilon} \omega/c}}{r} 
\frac{e^{-i d_1 \frac{\omega}{c}(\beta^{-1} - \sqrt{\varepsilon} \cos{\Theta})}- e^{i d_2 \frac{\omega}{c} (\beta^{-1} - \sqrt{\varepsilon} \cos{\Theta})}}{(\beta^{-1} - \sqrt{\varepsilon} \cos{\Theta})(1 - \beta^2 + (\beta \sqrt{\varepsilon} \sin{\Theta})^2)}\{\sin {\phi}, - \cos {\phi}, 0\} \cr \displaystyle && \qquad \times \Big ( \sin{\Theta} (1 - \beta^2 - \beta \sqrt{\varepsilon} \cos{\Theta})- a \frac{\omega}{v \gamma} \sin{\Theta} (1 - \beta^2 - \beta \sqrt{\varepsilon} \cos{\Theta}) J_0\Big (a \frac{\omega}{c} \sqrt{\varepsilon} \sin{\Theta}\Big ) 
K_1\Big (a \frac{\omega}{v \gamma}\Big ) + \cr && \displaystyle \qquad \qquad \qquad  + a \frac{\omega}{v \gamma^2} \sin{\Theta} (\cos{\Theta} + \beta \sqrt{\varepsilon} \sin^2{\Theta}) J_1\Big (a \frac{\omega}{c} \sqrt{\varepsilon} \sin{\Theta}\Big ) K_0\Big (a \frac{\omega}{v \gamma}\Big )  \Big ), \label{Eq3.5.9}
\end{eqnarray}
\end{widetext}
Then, following the procedure completely analogous to the one used in Sect.\ref{Sect3}, we consider refraction of the emitted radiation through the plane cylinder ``top'' 
laying in $xOy$ plane in Fig.\ref{Fig1}, that restricts us in what follows within the radiation angles $\theta \ll \pi/2$. Applying the reciprocity theorem, we come to the following formula 
for the energy radiated into vacuum (semi-space $z > 0$) from cylinder of length $d$ ($d_1 = d, d_2 = 0$):
\begin{widetext}
\begin{eqnarray}
\displaystyle && \frac{d^2W}{d\omega d\Omega} = \frac{e^2}{\pi^2 c}  \frac{\beta^2 \cos^2{\theta}}{(1 - \beta^2 \cos^2{\theta})^2} 
\Big | e^{-i d \frac{\omega}{c}\Big (\beta^{-1} - \sqrt{\varepsilon - \sin^2{\theta}}\Big )}- 1\Big |^2 
\Bigg |\frac{\varepsilon - 1}{(1 - \beta \sqrt{\varepsilon - \sin^2{\theta}})(\varepsilon \cos {\theta} + \sqrt{\varepsilon - \sin^2{\theta}})}
 \times \cr && \qquad \displaystyle \Big (\sin{\theta} (1 - \beta^2 - \beta \sqrt{\varepsilon - \sin^2{\theta}}) - a \frac {\omega}{v\gamma} \sin{\theta} (1 - \beta^2 - \beta \sqrt{\varepsilon - \sin^2{\theta}}) J_0\Big (a \frac{\omega}{c} \sin{\theta}\Big ) K_1\Big (a \frac{\omega}{v \gamma}\Big ) + \cr && \qquad \qquad \qquad \qquad \displaystyle + a \frac {\omega}{v\gamma^2}(\sqrt{\varepsilon - \sin^2{\theta}} + \beta \sin^2{\theta}) J_1\Big (a \frac{\omega}{c} \sin{\theta}\Big ) K_0\Big (a \frac{\omega}{v \gamma}\Big ) \Big )\Bigg |^2 \label{Eq3.5.10}
\end{eqnarray}
\end{widetext}
It is not difficult to see, that the limit $a \rightarrow \infty, d \rightarrow \infty$ leads again to the ordinary Ginzburg-Frank formula for forward TR (with contribution of CR) 
at the infinite boundary (\ref{Eq11}). Substitution of formulas for plasma permittivity (\ref{Eq3.5.1}) or (\ref{Eq3.5.8}) into (\ref{Eq3.5.10}) makes the last expression describing PR generated at the boundary of vacuum and cylinder of cold electron plasma. For the high density plasma and THz region of frequencies, one should use formula (\ref{Eq3.5.8}). For the high frequencies $\omega \gg \omega_{p(e)}$ (ultra-violet and X-ray regions), it is possible to use the simplest expression (\ref{Eq3.5.1}), see also \cite{T}. 
We emphasize, however, that for THz range formula (\ref{Eq3.5.8}) may be considered as just the first correction to (\ref{Eq3.5.1}) taking into account the possible local field effect 
in the high density plasma, but not taking into account the spatial dispersion. 
\begin{figure*}
\center \includegraphics[width=15.00cm, height=5.50cm]{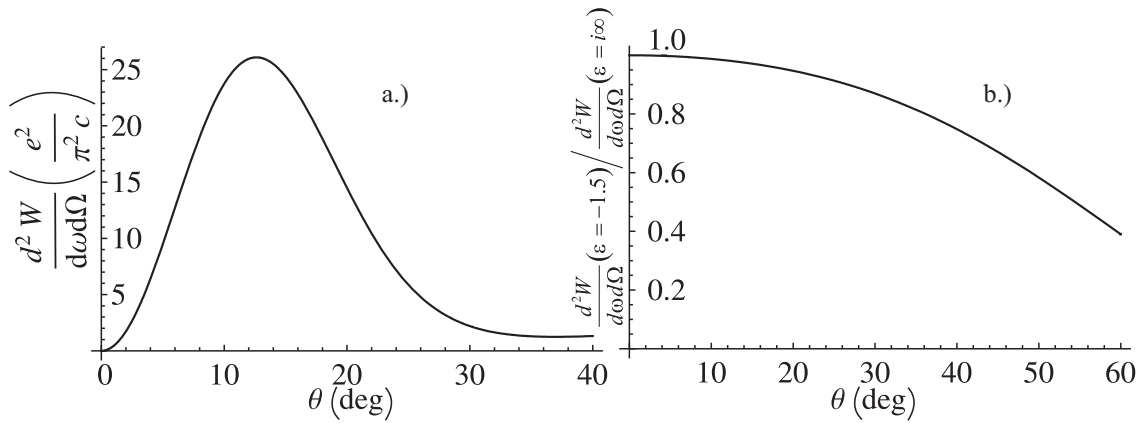}
\caption{\label{Fig6} Angular distributions of polarization radiation intensity in vacuum. Parameters: $\gamma = 50$, $\lambda = 100\mu$m, $a = 200\mu$m, $d = 1$mm, $\varepsilon^{\prime} = - 1.5$, $\varepsilon^{\prime \prime} = 0$.}
\end{figure*}

The angular distributions of the radiated energy are shown in Fig.\ref{Fig6}. The peak of TR is shifted to the large angles region due to small radius of cylinder $a \ll \gamma \lambda/(2 \pi)$. 
The negative value of permittivity results in the absence of CR, as the term $1 - \beta \sqrt{\varepsilon - \sin^2{\theta}}$ does not turn into zero. 
This feature is known to be the one of the radiation of an electron moving in the cold isotropic plasma, where only the longitudinal waves may be emitted \cite{Kuzelev}.
As is clear from Fig.\ref{Fig6} b.), the significant difference between the radiation characteristics for plasma with $\varepsilon = - 1.5$ and for the ideal conductor has a place in the region of large angles of emission $\theta \gg \gamma^{-1}$. But for the angles $\theta \sim \pi/2$ the model, strictly speaking, is not applicable.

If the radiation properties at large angles $\theta \sim \pi/2$ are of interest (for instance, in the non-relativistic case), one should consider refraction of the exact field (\ref{Eq3.5.9}) through the concave surface of cylinder (not through the back surface, like in (\ref{Eq3.5.10})). The last problem may be simplified in the two limit cases: 1.) $a \precsim \lambda$ and 2.) $a \gg \lambda$.
In the first case, the channel represents the small spatial inhomogeneity (see also the next Section), and applying of reciprocity theorem is not necessary.
In the second case, the surface of cylinder is almost flat for such waves, and refraction through this surface may be calculated as through the flat one.

It is worth noting the special case of partially ionized gas with some quantity of bound electrons possessing their own frequency $\omega_0$. 
If the damping constant $\Gamma_i$ in (\ref{Eq3.5.7}) is negligibly small, and the frequency of interest $\omega$ is close to $\omega_0$, 
the terms of bound electrons in (\ref{Eq3.5.7}) dominate:
\begin{eqnarray}
&& \displaystyle \varepsilon (\omega ) \approx 1 - \frac{\omega_{p(i)}^2}{\omega^2 + \displaystyle \frac{1}{3}\omega_{p(i)}^2 - \omega_0^2} \rightarrow -2, \label{Eq3.5.11}
\end{eqnarray}
when $\omega \rightarrow \omega_0$ (though neglect by the local field effect would led again to $- \infty$). Such a model allows to consider PR at the frequencies 
which are close to the natural ones of the medium. Note that DR on such a resonance frequency was considered for the first time in Ref.\cite{Ryaz-RF}.

\section{\label{Sect4} Smith-Purcell radiation as polarization radiation}

\begin{figure}
\center \includegraphics[width=8.00cm, height=5.00cm]{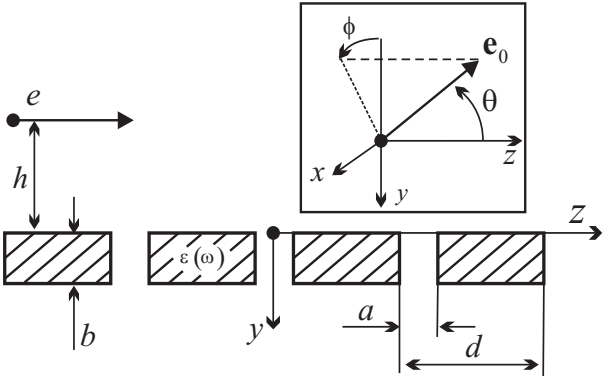}
\caption{\label{Fig7}Generation of polarization radiation by a charged particle moving nearby a grating.}
\end{figure}

Another type of polarization radiation is generated when a particle moves uniformly at the distance $h$ parallel to a periodic surface (grating) \cite{SPR}. 
The simplest target for this type of radiation is a set of $N$ rectangular $\infty \times b \times (d-a)$ strips made of material with permittivity $\varepsilon(\omega)$ and separated by vacuum gaps. 
Here $d$ is the grating period, $a$ is the vacuum gap width, see Fig.\ref{Fig7}. Physically, there is no difference between Smith-Purcell radiation and the ordinary diffraction radiation. Furthermore, they both (as well as TR, CR and PXR) arise because of polarization of the medium under the action of external field of the particle. And only periodicity of the grating changes the radiation spectrum, 
that makes SPR different from the ordinary DR.

Equality (\ref{Eq4}) represents the exact solution of Maxwell's equations describing all the types of PR inside a medium. But it was deduced only for a ``simply connected domain'', 
and in the general case it is not correct for the completely separated regions. It would be correct for a grating with continuously deformed profile, for example, rectangular or sinusoidal one. 
However, if the gap's width $a$ is much less than a wavelength in the transparent medium $a \ll \lambda/\sqrt{\varepsilon (\omega)}$, the gaps reveal themselves as the small spatial inhomogeneities (see also Ref.\cite{Zhevago}). The scattering cross-section of a plane wave on such ``particles'' is proportional to their polarizability \cite{L}, 
which is obviously equal to zero as $\varepsilon_{vac} = 1$. As regarding a conducting grating, in this case the waves are emitted only from the thin layer near the surface of a strip (skin depth), 
and they do not reach the interface as the transverse plane waves. Therefore, condition of applicability for Eq.(\ref{Eq4}) becomes just $a \ll \lambda$.

In order to find the energy radiated into vacuum, one should consider refraction of a spherical wave on the rectangular surface of a strip. The last problem may be simplified supposing that condition
$b \ll (d-a)$ is fulfilled, that excludes from the following consideration the angles of radiation which are too close to the surface ($\theta \rightarrow 0, \pi$, see similar calculations for a rectangular screen in Ref.\cite{JETPL}). In this case, it is possible to consider only refraction on the upper (or lower) plane of a strip.
Applying again the reciprocity theorem, one should find the field refracted from vacuum into medium through the surface of a strip. 
But according to the Smith-Purcell relation (see below), the last one has a dimension of order of a wavelength ($d \sim \lambda$), and therefore cannot be considered 
as the homogeneous flat surface. So, the only case allowing to apply the ordinary Fresnel laws of refraction is the geometry when the same condition $a \ll \lambda$ is fulfilled, 
and the surface of the grating is almost homogeneous for such waves (see e.g. \cite{Zhevago}). However, this condition does not bound the values of permittivity 
$\varepsilon (\omega) = \varepsilon^{\prime} + i \varepsilon^{\prime \prime}$ being considered, so it is possible to develop the exact theory of Smith-Purcell radiation 
in the sense of taking into account the real properties of the grating material.

The radiation field in the wave zone is found from Eq.(\ref{Eq4}) in the following way
\begin{eqnarray}
&& \displaystyle \bold H^{R} (\bold r, \omega ) = \frac{\omega}{2 c} (\varepsilon (\omega) - 1) \frac{e^{i r \sqrt{\varepsilon (\omega)} \omega/c}}{r} \times \cr && \ \bold k \times 
\int dz^{\prime} \int \limits_0^b dy^{\prime} \ \bold E^{0} (k_x, y^{\prime}, z^{\prime}, \omega) e^{-i k_y y^{\prime} - i k_z z^{\prime}}, \label{Eq4.1}
\end{eqnarray}
where the integral over $z^{\prime}$ represents a sum of the integrals over each strip of the grating.
The Fourier transform of the particle field now has the form (see e.g. \cite{J})
\begin{eqnarray}
&& \displaystyle \bold E^{0} (k_x, y^{\prime}, z^{\prime}, \omega) = \frac{- i e}{2 \pi v} \frac{e^{i \frac{\omega}{v}z^{\prime}}}{\sqrt{1 + \varepsilon (\beta \gamma e_x)^2}}
\{ \sqrt{\varepsilon}\beta\gamma e_x, \cr && \quad \displaystyle i \sqrt{1 + \varepsilon (\beta \gamma e_x)^2}, \gamma^{-1}\} e^{-(y^{\prime} + h) \frac{\omega}{v \gamma}
\sqrt{1 + \varepsilon (\beta \gamma e_x)^2}}, \label{Eq4.2}
\end{eqnarray}
Integration over $z^{\prime}$ gives the well-known product of interference terms:
\begin{eqnarray}
&& \displaystyle \int dz^{\prime} e^{i z^{\prime} (\frac{\omega}{v} - k_z) } = \displaystyle \frac{2 \sin{\Big (\frac{d - a}{2}(\frac{\omega}{v} - k_z)\Big )}}{\frac{\omega}{v} - k_z} \times
\cr && \qquad \qquad \displaystyle \frac{\sin{\Big (N \frac{d}{2}(\frac{\omega}{v} - k_z)\Big )}}{\sin{\Big (\frac{d}{2}(\frac{\omega}{v} - k_z)\Big )}} = F_{strip} F_N. \label{Eq4.3}
\end{eqnarray}
Substitution of (\ref{Eq4.2}) and (\ref{Eq4.3}) into (\ref{Eq4.1}) leads to the following expression for radiation field (dependence of $\varepsilon$ on $\omega$ below is implied):
\begin{widetext}
\begin{eqnarray}
\displaystyle && \bold H^{R} = \frac{i e \omega^2}{2 \pi v c^2} \sqrt{\varepsilon} (\varepsilon - 1) \frac{e^{i r \sqrt{\varepsilon} \frac{\omega}{c}}}{r}
\Big \{\gamma^{-1} e_y - i e_z \sqrt{1 + \varepsilon (\beta \gamma e_x)^2}, e_x ( \sqrt{\varepsilon} \beta \gamma e_z - \gamma^{-1}),  \displaystyle e_x (i \sqrt{1 + \varepsilon (\beta \gamma e_x)^2} - \sqrt{\varepsilon} \beta \gamma e_y) \Big \}\cr && \qquad \times \frac{\exp{\Big (- b \Big (i \frac{\omega}{c} \sqrt{\varepsilon} e_y + \frac{\omega}{v \gamma}\sqrt{1 + \varepsilon (\beta \gamma e_x)^2}\Big )\Big )} - 1}{\sqrt{1 + \varepsilon (\beta \gamma e_x)^2} (i \frac{\omega}{c} \sqrt{\varepsilon} e_y + \frac{\omega}{v \gamma}\sqrt{1 + \varepsilon (\beta \gamma e_x)^2})}
\frac{\sin{\Big (\frac{d - a}{2}(\frac{\omega}{v} - k_z)\Big )}}{\frac{\omega}{v} - k_z} \displaystyle \frac{\sin{\Big (N \frac{d}{2}(\frac{\omega}{v} - k_z)\Big )}}{\sin{\Big (\frac{d}{2}(\frac{\omega}{v} - k_z)\Big )}} e^{-h \frac{\omega}{v \gamma}\sqrt{1 + \varepsilon (\beta \gamma e_x)^2}}, \label{Eq4.4}
\end{eqnarray}
\end{widetext}

If the particle moves above a homogeneous medium ($N = 1,\ d - a \rightarrow \infty$) the interference term is transformed into delta-function:
\begin{eqnarray}
\displaystyle && \frac{\sin{\Big (\frac{d - a}{2}(\frac{\omega}{v} - k_z)\Big )}}{\frac{\omega}{v} - k_z}\rightarrow \pi \delta \Big (\frac{\omega}{v} - k_z \Big ), \label{Eq4.5}
\end{eqnarray}
determining the Cherenkov condition. However, this field propagates inside the medium and is not radiated into vacuum through the upper (lower) surface (see below and also \cite{Bol}).

The next step is to find the energy radiated into vacuum using the reciprocity theorem. As we already used the condition $a \ll \lambda$, one may consider the grating as the homogeneous interface
and apply the same rules of refraction as in the previous paragraphs. As the problem being solved has no azimuthal symmetry, the final expression for the absolute value of the radiation field in vacuum 
becomes (see also \cite{JETPL}):
\begin{eqnarray}
\displaystyle && |\bold E^{R}_{vac}|^2 = T_{\perp} |H^R_{\perp}|^2 + T_{\parallel}(|H_y^R|^2 + |H_{\parallel}^R|^2), \label{Eq4.6}
\end{eqnarray}
where
\begin{eqnarray}
\displaystyle && H^R_{\perp} = H^R_x \cos \tilde \phi - H^R_z \sin \tilde \phi, \cr && H^R_{\parallel} = H^R_x \sin \tilde \phi + H^R_z \cos \tilde \phi \label{Eq4.7}
\end{eqnarray}
are the components of magnetic field (\ref{Eq4.4}) perpendicular and parallel to the plane of incidence of the wave on the interface of the grating, and also:
\begin{eqnarray}
\displaystyle && T_{\perp} = \Big |\frac{f_H}{\varepsilon}\Big |^2 = \Bigg |\frac{2 \sin \theta \cos \phi}{\varepsilon \sin \theta \cos \phi + \sqrt{\varepsilon - 1 + (\sin \theta \cos \phi)^2}}\Bigg |^2, \cr 
\displaystyle && T_{\parallel} = \Big |\frac{f_E}{\sqrt{\varepsilon}}\Big |^2 = \cr \displaystyle && \qquad = \Bigg |\frac{2 \sin \theta \cos \phi} {\sqrt{\varepsilon} (\sin \theta \cos \phi + \sqrt{\varepsilon - 1 + (\sin \theta \cos \phi )^2})}\Bigg |^2 \label{Eq4.8}
\end{eqnarray}
are the transmission coefficients expressed through the vacuum angular variables used in the Smith-Purcell radiation problems:
\begin{eqnarray}
\displaystyle && \bold e_0 = \{\sin \theta \sin \phi, - \sin \theta \cos \phi, \cos \theta\}, \label{Eq4.9}
\end{eqnarray}
and $f_H, f_E$ are the Fresnel coefficients \cite{L}. As may be seen in Fig.\ref{Fig6}, the azimuth angle $\phi$ is counted from the negative direction of $y$-axis, because we consider radiation emitted into the upper semi-space $y < 0$. The azimuth angle $\tilde \phi$ in formula (\ref{Eq4.7}) belongs to the plane of interface: $xOz$. Connection of this angle with variables (\ref{Eq4.9}) 
is found by equating the components of (\ref{Eq4.9}) and the ones written through the angles used in a problem of refraction (see e.g. \cite{L}). The final result is:
\begin{eqnarray}
\displaystyle \sin \tilde \phi = \frac{\sin \theta \sin \phi}{\sqrt{1 - (\sin \theta \cos \phi)^2}}, \cos \tilde \phi = \frac{\cos \theta}{\sqrt{1 - (\sin \theta \cos \phi)^2}}. \label{Eq4.10}
\end{eqnarray}

The last step is to express the unit vector of radiation in the medium $\bold e = \bold r/r$ in Eq.(\ref{Eq4.4}) through the vacuum variables according to the Snell's law \cite{L}:
\begin{eqnarray}
\displaystyle \bold e = \frac{1}{\sqrt{\varepsilon}}\{\sin \theta \sin \phi, - \sqrt{\varepsilon - 1 + (\sin \theta \cos \phi)^2}, \cos \theta\}. \label{Eq4.11}
\end{eqnarray}
The minus before the square root is chosen so that the wave (\ref{Eq4.4}) is damped deep into absorbing medium.
Returning to the delta-function determining the condition of Cherenkov radiation (\ref{Eq4.5}), one may note that its argument written in the vacuum variables is proportional 
to $1/\beta - \cos \theta$ and never turns into zero. It means the absence of CR transmitted through the surface parallel to the particle trajectory, as we already indicated in Sect.\ref{Sect3}.

Now we can write down the final expression for the energy radiated in the upper semi-space:
\begin{widetext}
\begin{eqnarray}
&& \displaystyle\frac{d^2 W}{d \omega d \Omega} = \frac{e^2}{\pi^2 c} \frac{(\beta \sin \theta \cos \phi)^2}{(1 - \beta^2 + (\beta \sin \theta \sin \phi)^2) (1 - (\sin \theta \cos \phi)^2)}\frac{\sin^2{\Big (\frac{d - a}{2} \frac{\omega}{c}(\frac{1}{\beta} - \cos \theta)\Big )}}{(1 - \beta \cos \theta)^2}
\frac{\sin^2{\Big (N \frac{d}{2}\frac{\omega}{c}(\frac{1}{\beta} - \cos \theta)\Big )}}{\sin^2{\Big (\frac{d}{2}\frac{\omega}{c}(\frac{1}{\beta} - \cos \theta)\Big )}}\times \cr && \displaystyle \Bigg |\displaystyle \frac{(\varepsilon - 1) (\exp\{- b \frac{\omega}{v \gamma} (\sqrt{1 + (\beta \gamma \sin \theta \sin \phi )^2} - i \beta \gamma \sqrt{\varepsilon - 1 + (\sin \theta \cos \phi )^2})\} - 1)}{\sqrt{1 + (\beta \gamma \sin \theta \sin \phi )^2} - i \beta \gamma \sqrt{\varepsilon - 1 + (\sin \theta \cos \phi )^2}} \Bigg |^2 \times \cr && \displaystyle \Bigg (\Big |\frac{(\gamma^{-1} \cos \theta + \beta \gamma (\sin \theta \sin \phi)^2) \sqrt {\varepsilon - 1 + (\sin \theta \cos \phi)^2} + i (1 - (\sin \theta \cos \phi)^2) \sqrt{1 + (\beta \gamma \sin \theta \sin \phi)^2}}{\varepsilon \sin \theta \cos \phi + \sqrt {\varepsilon - 1 + (\sin \theta \cos \phi)^2}}\Big |^2 \cr && \displaystyle + (\sin \theta \sin \phi)^2 (\gamma^{-1} - \beta \gamma \cos \theta)^2 \frac{1 - (\sin \theta \cos \phi)^2 + |\sqrt{\varepsilon - 1 + (\sin \theta \cos \phi )^2}|^2}{|\sqrt{\varepsilon} (\sin \theta \cos \phi + \sqrt{\varepsilon - 1 + (\sin \theta \cos \phi )^2})|^2} \Bigg ) e^{-h \frac{2\omega}{v \gamma}\sqrt{1 + (\beta \gamma \sin \theta \sin \phi)^2}}. \label{Eq4.12}
\end{eqnarray}
\end{widetext}
\begin{figure*}
\center \includegraphics[width=15.00cm, height=5.50cm]{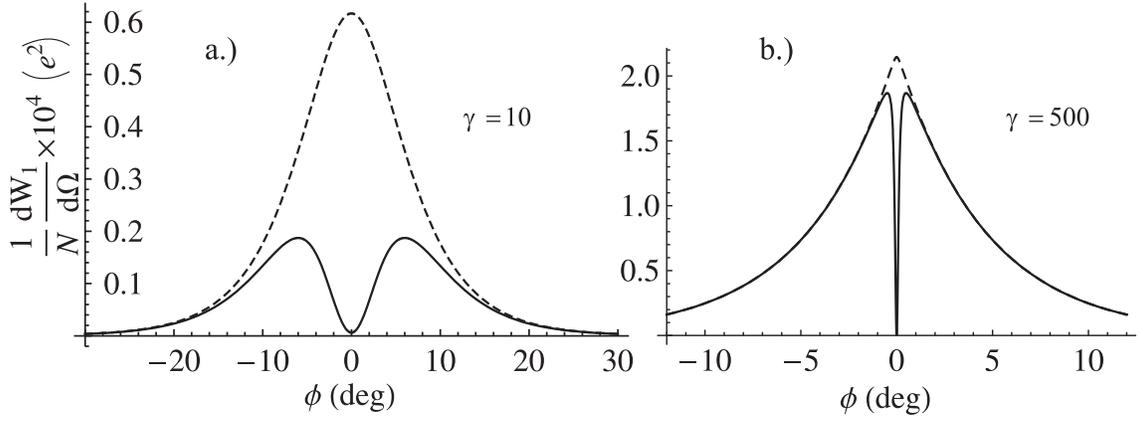}
\caption{\label{Fig8} Comparison of SPR angular distributions for the ideally conducting grating according to formula (\ref{Eq4.16-2}) - solid line, and according to the model \cite{Br} - dashed line. 
Parameters: $\theta = 90^\circ$, $d = 1$mm, $a = 0.2$mm, $h = 1$mm, $m = 1$.}
\end{figure*}

If the number of strips is large enough ($N \gg 1$) the interference term squared gives another delta-function:
\begin{eqnarray}
\displaystyle && \frac{\sin^2{\Big (N \frac{d}{2}\frac{\omega}{c}(\frac{1}{\beta} - \cos \theta)\Big )}}{\sin^2{\Big (\frac{d}{2}\frac{\omega}{c}(\frac{1}{\beta} - \cos \theta)\Big )}} \rightarrow \cr && \qquad \rightarrow 2 \pi N \sum \limits_{m = 1}^{\infty}\delta \Big (d \frac{\omega}{c}\Big (\frac{1}{\beta} - \cos \theta \Big ) - 2 \pi m \Big ). \label{Eq4.13}
\end{eqnarray}
Its zeros give the well-known Smith-Purcell relation \cite{SPR}:
\begin{eqnarray}
\displaystyle && \lambda_m = \frac{d}{m} \Big (\frac{1}{\beta} - \cos \theta \Big ). \label{Eq4.14}
\end{eqnarray}

Although the formula (\ref{Eq4.12}) is rather complicated, it permits to do the following factorization
\begin{eqnarray}
\displaystyle && \frac{d^2 W}{d \omega d \Omega} = \frac{d^2 W}{d \omega d \Omega}\Big|_{sc} F^2_{strip} F^2_N, \label{Eq4.15}
\end{eqnarray}
where $d^2 W/d \omega d \Omega|_{sc}$ is the energy radiated while a charge moves above a semi-infinite screen. In other words, the first factor in (\ref{Eq4.15})
describes DR generated on a semi-infinite screen of the finite permittivity (conductivity) and height. The second factor describes interference of the waves emitted from the left and right sides 
of a strip. Finally, the third factor takes into account the interference of waves emitted from all $N$ periods of the grating.
This factorization is known in literature for the case of an ideally conducting grating \cite{PRE}. 
In the present theory, it appears as a result of condition $a \ll \lambda$ used when deriving Eq.(\ref{Eq4.12}). 
In fact, this limitation is quite general, because if the opposite condition $a \gg \lambda$ holds true, it is impossible to consider radiation from a single strip (even the ideally conducting one) 
to be entirely independent of radiation from the other one. In this case, the secondary re-reflections of radiation may be of importance, and the formula for the radiated energy would not have the structure (\ref{Eq4.15}). Note that such factorization does not have a place also in the van den Berg's model of SPR (calculations within this model for the grating being considered 
were performed in Ref.\cite{Kube}). It is so, because in this model the limitation $b \ll (d-a)$ was not used. Unfortunately, it was done only for the ideally conducting gratings.

To begin analysis of the formula (\ref{Eq4.12}), let us consider the simplest and the most investigated case of ideal conductivity: $\varepsilon^{\prime \prime} \rightarrow \infty$.
Performing the corresponding transition, we have
\begin{widetext}
\begin{eqnarray}
&& \displaystyle \frac{d^2 W}{d \omega d \Omega} \rightarrow \frac{e^2}{\pi^2 c} \frac{\gamma^{-2} (1 - (\sin \theta \sin \phi )^2 ) + 2 \beta \cos \theta (\sin \theta \sin \phi )^2
+ (\beta \gamma )^2 \sin^4 \theta \sin^2 \phi}{1 + (\beta \gamma \sin \theta \sin \phi)^2} \times \cr && \qquad \qquad \qquad \qquad \qquad \qquad \displaystyle
\frac{\sin^2{\Big (\frac{d - a}{2} \frac{\omega}{c}(\frac{1}{\beta} - \cos \theta)\Big )}}{(1 - \beta \cos \theta)^2}
\frac{\sin^2{\Big (N \frac{d}{2}\frac{\omega}{c}(\frac{1}{\beta} - \cos \theta)\Big )}}{\sin^2{\Big (\frac{d}{2}\frac{\omega}{c}(\frac{1}{\beta} - \cos \theta)\Big )}}
e^{-h \frac{2\omega}{v \gamma}\sqrt{1 + (\beta \gamma \sin \theta \sin \phi)^2}}, \label{Eq4.16}
\end{eqnarray}
\end{widetext}
where dependence upon the grating height $b$ has disappeared due to the skin-effect on the upper facet of each strip.
For the large number of strips, we can integrate this expression over frequency using (\ref{Eq4.13}). The result for radiation on the m-th harmonics is
\begin{eqnarray}
&& \displaystyle \frac{1}{N}\frac{d W_m}{d \Omega} = \frac{2 e^2 \beta}{\pi d} R(\gamma, \theta, \phi ) 
\frac{\sin^2{\Big (\pi m \frac{d - a}{d}\Big )}}{(1 - \beta \cos \theta)^3} \times \cr && \displaystyle \qquad \qquad \qquad e^{-h \frac{4 \pi}{\beta \gamma \lambda_m}\sqrt{1 + (\beta \gamma \sin \theta \sin \phi)^2}}, \label{Eq4.16-2}
\end{eqnarray}
where we use denotation $R(\gamma, \theta, \phi )$ for angular part (the first line) of formula (\ref{Eq4.16}).

It is essentially that the formula (\ref{Eq4.16}) may be derived using completely different way. As we already noted, there exist several models of SPR for the ideally conducting gratings 
based on the presentation of SPR as a field of a surface current density induced on the grating by the field of the particle. 
But the common drawback of these models consists in the neglect of the normal to the interface component of the surface current (see details in \cite{PLA}).
Without taking into account this component, it is impossible to obtain even the simplest Ginzburg-Frank formula for intensity of transition radiation on an ideally conducting plane.
On the contrary, it is the exact expression for the surface current density found in Ref.\cite{PLA} 
\begin{eqnarray}
&& \displaystyle \bold {j}_{surf} \simeq \frac{c}{2 \pi}\ \bold {e}_0 \times \Big [\bold n \times \bold {E}^0 \Big ] \label{Eq4.17}
\end{eqnarray}
that leads to the well-known results for TR (including geometry of oblique incidence). Here, $\bold n$ is the normal vector to the surface.
This expression allows to obtain exact formulas for DR, and in the special case of the ideally conducting grating it leads to the very same formula (\ref{Eq4.16}) for Smith-Purcell radiation \cite{PHD}.
Lastly, the part $d^2 W/d \omega d \Omega|_{sc}$ of (\ref{Eq4.16}) corresponding to DR from a semi-plane completely coincides with the one derived in Refs.\cite{PLA, J, JETPL}.
\begin{figure*}
\center \includegraphics[width=17.00cm, height=5.50cm]{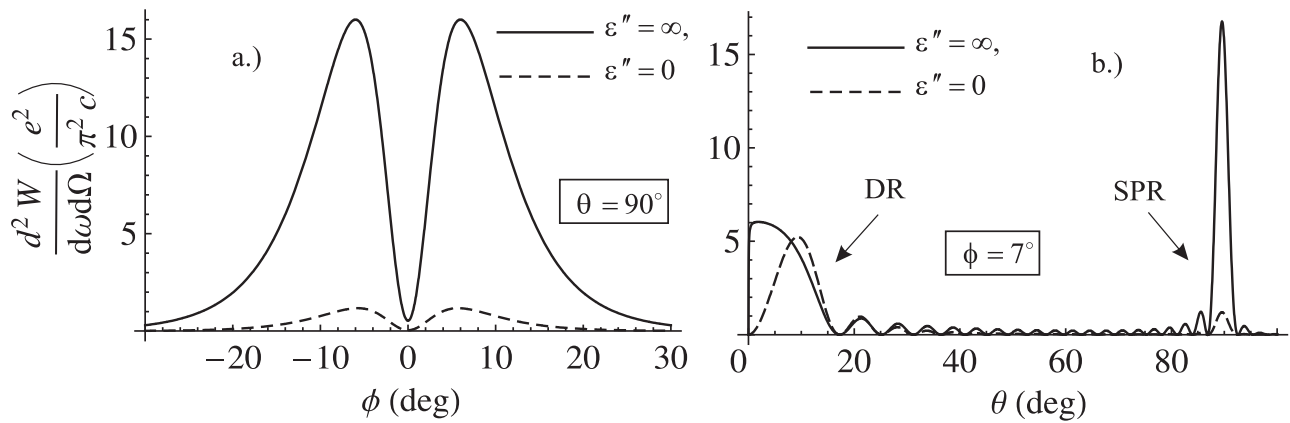}
\caption{\label{Fig9} Angular distributions of the radiated energy. Parameters: $\gamma = 10$, $\varepsilon^{\prime} = 1.5$, $\lambda = 2$mm, $d = 2$mm, $a = 0.5$mm, $b = 0.5$mm, $h = 2$mm,
$N = 20$.}
\end{figure*}

Comparison of SPR angular characteristics according to formula (\ref{Eq4.16-2}) and the ones within the known surface current model \cite{Br} without the normal component 
is presented in Fig.\ref{Fig8}. Note that analog of formula (\ref{Eq4.16-2}) in the model \cite{Br} is obtained by the substitution: $R(\gamma, \theta, \phi) \rightarrow (\beta \sin \theta)^2$. 
It is easy to see, that in addition to disappearance of the important angular terms corresponding to DR from a semi-plane, the extra $\beta^2$ arises. 
Completely the same excess term also has a place in the formula for TR on an ideally conducting plane obtained with the method of the surface current without normal component similar 
to that of Ref.\cite{Br} (see in more detail \cite{J, PLA}). However, in the ultrarelativistic limit ($\gamma \rightarrow \infty$), the function $R(\gamma, \theta, \phi)$ in (\ref{Eq4.16-2}) 
tends to $\sin^2 \theta$, and the minimum in the azimuthal dependence disappears (compare a.) and b.) in Fig.\ref{Fig8}). Thus, for ultrarelativistic energies the use of the surface current model \cite{Br} is justified (for example, for 28.5 GeV electrons of the experiment \cite{Doucas} at SLAC). For moderate relativistic electrons (hundreds of MeV and less), one should use the exact expression (\ref{Eq4.16-2}). Coincidence of both models in the ultrarelativistic limit is no surprise, because the model of a surface current without the normal component originates in the theory of diffraction of the ordinary plane waves \cite{J, PLA}. In the ultrarelativistic case, the field of an electron is almost transverse, that is why the methods of optics are quite applicable. 
We emphasize, however, the limitation of all the surface current models for the gratings with vacuum gaps: the gap width must be less than the wavelength studied.

For the opposite case of transparent material of the grating ($\varepsilon^{\prime \prime} = 0$), dependence on the grating height becomes
\begin{widetext}
\begin{eqnarray}
\displaystyle && \Big |\exp \Big \{- b \frac{\omega}{v \gamma} \Big (\sqrt{1 + (\beta \gamma \sin \theta \sin \phi )^2} - i \beta \gamma \sqrt{\varepsilon - 1 + (\sin \theta \cos \phi )^2}\Big )\Big \} - 1 \Big |^2 = 
\cr \displaystyle && \qquad = 4 \Big (\sinh^2\Big (\frac{b}{2}\frac{\omega}{v \gamma} \sqrt{1 + (\beta \gamma \sin \theta \sin \phi )^2}\Big ) + \sin^2 \Big (\frac{b}{2}\frac{\omega}{c} \sqrt{\varepsilon - 1 + (\sin \theta \cos \phi )^2} \Big ) \Big ) e^{-b \frac{\omega}{v \gamma} \sqrt{1 + (\beta \gamma \sin \theta \sin \phi )^2}}. \label{Eq4.18}
\end{eqnarray}
\end{widetext}
For the values of $b \ll \beta \gamma \lambda/\pi$ where $\sinh^2 {(x)} \approx x^2 \ll 1$, this implies the almost periodical dependence of the radiation intensity as on the grating height $b$, 
as on the value of permittivity $\varepsilon \equiv \varepsilon^{\prime}$. However, for conducting material of the grating with $\varepsilon^{\prime \prime} \gg 1$, 
the radiated energy weakly depends upon the value of permittivity.

In the general case with $\varepsilon (\omega) = \varepsilon^{\prime} + i \varepsilon^{\prime \prime}$, intensity of Smith-Purcell radiation for a conducting grating several times exceeds 
the one for the grating of transparent material, see Fig.\ref{Fig9}. On the contrary, intensity of the ordinary DR generated within small polar angles weakly depends on the value of permittivity. 
This \textit{forward} diffraction radiation is generated on the grating as on a semi-infinite screen.

It should be noted, that the well-known absence of radiation on even harmonics for $a = d/2$ predicted by the surface current models of SPR (see (\ref{Eq4.16-2}) and \cite{Br, PRE}) 
appears just due to the fact, that all they are applicable only if condition $a \ll \lambda \sim d$ is fulfilled. Argumentation presented in the beginning of this Section clearly demonstrates the origin of this restriction. In conclusion of the Section, we emphasize that solution for SPR derived above represents the exact solution of the problem under the conditions used: small vacuum gaps, thin strips. In order to find an analogous solution for a grating with continuously deformed profile (rectangular or sinusoidal one) it is not necessary to use the restriction of the small gaps. But in order to overcome condition $b \ll (d-a)$, we have to know the laws of refraction of a plane wave on a wedge having the right opening angle and made of material with $\varepsilon = \varepsilon^{\prime} + i \varepsilon^{\prime \prime}$. Though we do not know the solution of such a problem, it may be useful to apply the reciprocity theorem at least for a conducting wedge using the well-known Sommerfeld's solution (see, for example, \cite{L}).

\section{\label{Sect6} Conclusion}

As is shown in this paper, the well-known formulas for Cherenkov radiation, Transition radiation and Diffraction radiation generated by the charged particles in the targets of the real dielectric properties 
may be obtained by solving Maxwell's equations with polarization current density (\ref{Eq1}) in the right-hand side. Such universal description proves the physical equivalence of the various types of polarization radiation. Moreover, this approach allows to calculate with a good accuracy characteristics of TR, DR, SPR and CR generated in the targets of the given shape and finite permittivity $\varepsilon = \varepsilon^{\prime} + \varepsilon^{\prime \prime}$, that is of relevance for development of new techniques of beam diagnostics and new types of radiation sources. 
In particular, the solution found for SPR problem is applicable for the arbitrary energies of particles (in contrast, for example, to the model \cite{Br}) and the arbitrary value of the grating's permittivity. 
The model developed also allows to simulate the radiation characteristics from an electron beam transported through the special dielectric structure, for instance, a cylindrical waveguide 
with a grating produced on the internal surface of the one. In this case, radiation will be emitted through external surface of the cylinder and may be detected without problems. 
Such a scheme provides minimal beam perturbation because of the azimuthal symmetry and low SPR energy losses, and it may be used for non-destructive beam diagnostics.
Finally, the exact taking into account of the spatial dispersion of the target's permittivity may be also performed, that allows to develop more rigorous models, for example, 
for PR in the spatially inhomogeneous plasma.

It should be mentioned that in order to detect CR in vacuum from a target with $\varepsilon^{\prime} \gg 1$ (i.e. when condition $\sin^2 \theta = |\varepsilon - 1/\beta^{2}| < 1$
is not fulfilled), the normal vector to the ``exit'' surface should coincide with CR direction in the medium $\Theta = \arccos {(1/(\beta \sqrt{\varepsilon}))}$. 
Calculations for a target made of material possessing weak dispersion and having such inclined surface may be performed using directly Eq.(\ref{Eq4}) and neglecting by refraction at the boundary.
It allows to develop the new technique for the non-intercepting bunch length diagnostics based on the Cherenkov radiation generated in the targets of the special shape \cite{China}.

In conclusion, we note that the theory presented is developed for the simplest case of PR generated by the uniformly moving point charge. 
But it may be comparatively easy generalized for radiation of $N$ non-interacting point charges, that allows to take into account possible \textit{coherence effects} for all the PR types.
Finally, the external field $\bold {E}^0$ may correspond to the field of a multipole (multipoles) or even of a non-uniformly moving particle (particles), 
that is of importance, for example, for modern laser-plasma based THz radiation sources \cite{Leemans}.

\begin{acknowledgments}
We are grateful to M.I. Ryazanov, A.A. Tishchenko and L.G. Sukhikh for fruitful discussions.
This work is partially supported by Russian Federal Agency for Education under contracts No. $\Pi$ 1143 and No. $\Pi$ 617, 
and Russian Science and Innovations Federal Agency under contract No. 02.740.11.0238.
\end{acknowledgments}

\

\appendix*

\section{Exact solution of Maxwell's equations}

In the paper we considered the radiation processes related to the far-field part of polarization current density field. 
But along with transverse waves of polarization radiation, there may also exist longitudinal waves corresponding to the zeros of permittivity of the target material.
The possibility of existence for such waves is clear at least from the fact, that when deriving far-field equality (\ref{Eq5}) from the general expression (\ref{Eq4}),
we used the condition $r \gg \lambda/\sqrt{\varepsilon (\omega)}$. This condition cannot be satisfied if $\varepsilon (\omega) \rightarrow 0$, 
and the ordinary wave zone formulas used above become inapplicable. On the other hand, at the small enough distances (actually, if $r \precsim \gamma \lambda/(2 \pi)$) 
longitudinal fields may be of importance, even if the value of permittivity is not close to zero. Such fields are often called \textit{wake fields}, 
but the theoretical methods used in the corresponding theories noticeably differ from those used in the theories of polarization radiation.
In particular, the physical analogy between the wake fields theory and the one of TR and DR was demonstrated relatively recently and only for some simplest cases 
of the ideally conducting targets and ultrarelativistic energies of the particles (see, for example, \cite{B, Stupakov, X, Xiang}).

To demonstrate the origin of longitudinal and transverse waves, it is necessary to derive the general expressions for electric field strength and magnetic field strength 
of polarization current density. As we already know the exact expression for magnetic field inside the medium (\ref{Eq4}),
we can find the analogous formula for electric field using Maxwell's equations. The simplest way is to use the one:
\begin{eqnarray}
\displaystyle && \curl \bold H^{pol}(\bold r, \omega ) + i \frac{\omega}{c} \bold E^{pol}(\bold r, \omega ) = \frac{4 \pi}{c} \bold j^{pol}(\bold r, \omega ). \label{Eq5-1}
\end{eqnarray}
And thus
\begin{eqnarray}
\displaystyle \bold E^{pol}(\bold r, \omega ) = \frac{i c}{\omega \varepsilon (\omega )} \Big (\curl \bold H^{pol} - \frac{4 \pi}{c} \bold j^{pol (0)}\Big ) , \label{Eq5-2}
\end{eqnarray}
where $\bold j^{pol (0)} = \sigma (\omega) \bold E^0 (\bold r, \omega )$ is denoted. 
As the point of observation $\bold r$ is disposed inside the homogeneous medium (target of a given shape), $\varepsilon$ and $\sigma$ are independent of coordinates.
Then the general expression for magnetic field suitable at the arbitrary distances is found from (\ref{Eq4}) as
\begin{eqnarray}
\displaystyle && \bold H^{pol}(\bold r, \omega ) =  \frac{1}{c} \int \limits_{V_T} \frac{\bold r - \bold r^{\prime}}{|\bold r - \bold r^{\prime}|} \times \bold j^{pol (0)}(\bold r^{\prime}, \omega )
\cr \displaystyle && \ \times \Big (i \sqrt{\varepsilon (\omega)} \frac{\omega}{c} - \frac{1}{|\bold r - \bold r^{\prime}|} \Big ) \frac{e^{i \sqrt{\varepsilon (\omega )} \omega |\bold r - {\bold r}^{\prime}|/c}}{|\bold r - {\bold r}^{\prime}|} d^3 r^{\prime}. \label{Eq5-3}
\end{eqnarray}
Taking $\curl$ of this expression according to (\ref{Eq5-2}), it is possible to derive the following formula for electric field strength: 
\begin{widetext}
\begin{eqnarray}
\displaystyle && \bold E^{pol}(\bold r, \omega ) = \frac{i}{\omega \varepsilon (\omega )} \int \limits_{V_T} \Big \{\bold {j}^{pol (0)} (\bold r^{\prime}, \omega ) \ \Big ( \varepsilon(\omega) \Big (\frac{\omega}{c}\Big )^2 + \frac{i \sqrt{\varepsilon(\omega)} \omega/c}{|\bold r - \bold r^{\prime}|}  - \frac{1}{|\bold r - \bold r^{\prime}|^2} \Big ) 
- \frac{\bold r - \bold r^{\prime}}{|\bold r - \bold r^{\prime}|} \Big (\frac{\bold r - \bold r^{\prime}}{|\bold r - \bold r^{\prime}|}, \bold {j}^{pol (0)}\Big ) \cr \displaystyle && \qquad \displaystyle \times \Big (\varepsilon(\omega) \Big (\frac{\omega}{c}\Big )^2+ \frac{3 i \sqrt{\varepsilon(\omega)} \omega/c}{|\bold r - \bold r^{\prime}|}  - \frac{3}{|\bold r - \bold r^{\prime}|^2} \Big ) \Big \} 
\frac{e^{i\sqrt{\varepsilon (\omega )}\omega|\bold r - \bold r^{\prime}|/c}}{|\bold r - \bold r^{\prime}|}d^3 r^{\prime} + \Big (\frac{1}{\varepsilon (\omega)} - 1 \Big ) \bold E^0 (\bold r, \omega ). \label{Eq5-4}
\end{eqnarray}
\end{widetext}
where we used relation (\ref{Eq8.5}) between conductivity and permittivity, and integration is performed over the target volume. 
It is the second term in Eq.(\ref{Eq5-4}) 
\begin{eqnarray}
\displaystyle && \bold E^{w} \equiv - i \frac{4 \pi} {\omega \varepsilon (\omega)} \ \bold j^{pol (0)} = \Big (\frac{1}{\varepsilon (\omega)} - 1\Big ) \bold E^0, \label{Eq5-5}
\end{eqnarray}
that describes the longitudinal oscillations. Necessity of presence of this term may also be shown via another Maxwell's equation:
\begin{eqnarray}
\displaystyle && \div \bold E^{pol} = 4 \pi \rho^{pol}. \label{Eq5-6}
\end{eqnarray}
The polarization charge density $\rho^{pol}$ is related to the current density with continuity equation: 
\begin{eqnarray}
\displaystyle && \rho^{pol} = \frac{1}{i \omega} \div \bold j^{pol} = \frac{\sigma (\omega)}{i \omega}
(\div \bold E^{0} + \div \bold E^{pol}).\label{Eq5-7}
\end{eqnarray}
Comparing two last equations, we come to the following conclusion
\begin{eqnarray}
\displaystyle && \div \bold E^{pol} = 4 \pi \rho^{pol} = \div \Big (\Big ( \frac{1}{\varepsilon (\omega)} - 1\Big ) \bold E^0 \Big ). \label{Eq5-8}
\end{eqnarray}
As the integral term in (\ref{Eq5-4}) represents the curl of $\bold H^{pol}$, divergence of (\ref{Eq5-4}) completely coincides with (\ref{Eq5-8}).
The charge density $\rho^{w}\equiv \rho^{pol}$ is sometimes called the density of a wake charge \cite{Ryaz}. 
The field (\ref{Eq5-5}) $\bold E^{w}$ describes the longitudinal waves produced inside the medium by the field of the charged particle.
Condition $\varepsilon (\omega) \rightarrow 0$ is known to be the one of appearance of such oscillations \cite{L, Ryaz}.

If the condition $|\bold r - \bold r^{\prime}| \gg \lambda/\sqrt{\varepsilon (\omega)}$ is fulfilled, formula (\ref{Eq5-4}) transforms into
\begin{eqnarray}
\displaystyle && \bold E^{pol}(\bold r, \omega ) = -\frac{i \omega}{c^2} \int \limits_{V_T} \frac{\bold r - \bold r^{\prime}}{|\bold r - \bold r^{\prime}|} \times \Big [\frac{\bold r - \bold r^{\prime}}{|\bold r - \bold r^{\prime}|} \times \bold {j}^{pol (0)} \Big ] \cr \displaystyle && \times \frac{e^{i\sqrt{\varepsilon (\omega )}\omega|\bold r - \bold r^{\prime}|/c}}{|\bold r - \bold r^{\prime}|}d^3 r^{\prime} + \Big (\frac{1}{\varepsilon (\omega)} - 1 \Big ) \bold E^0 (\bold r, \omega ). \label{Eq5-9}
\end{eqnarray}
The field of the uniformly moving charge in vacuum $\bold E^{0} (\bold r, \omega)$ decays exponentially (see e.g. \cite{J}). In an unbounded homogeneous medium, 
the effective region giving a contribution to the integral in (\ref{Eq5-4}) is $r_{eff} \simeq \gamma \lambda/(2 \pi)$. At the distances $r \gg \gamma \lambda/(2 \pi)$
the wake field may be ignored, and the formula (\ref{Eq5-9}) describes just transverse waves of polarization radiation:
\begin{eqnarray}
\displaystyle && \bold E^{pol} \approx \bold E^{R} = -\frac{i \omega}{c^2} \frac{1}{r} \ \bold e \times \bold e \times \cr \displaystyle 
&& \qquad \times \int \limits_{V_T} \bold {j}^{pol (0)} \exp\{i\sqrt{\varepsilon (\omega )}\omega |\bold r - \bold r^{\prime}|/c \} d^3 r^{\prime},
\cr \displaystyle && \bold H^{pol} \approx \bold H^{R} = \sqrt{\varepsilon (\omega)} \ \bold e \times \bold E^{R}, \label{Eq5-10}
\end{eqnarray}
where $\bold e = \bold r/r$ is denoted. As was shown in the previous Sections, these simple expressions describe all the types of polarization radiation inside the medium.
The expansion of exponent into $r^{\prime}/r \ll 1$ series leads to the formulas for polarization radiation in the wave zone ($r \gg \gamma^2 \lambda/(2 \pi)$) 
or in the so-called pre-wave zone ($\gamma^2 \lambda/(2 \pi) > r \gg \gamma \lambda/ (2 \pi)$). Note that in this paper we do not consider the possible interference between the fields 
of polarization currents (\ref{Eq5-3}), (\ref{Eq5-4}) and the ones of the moving particles (formation zone effects).

In another limiting case of small distances $|\bold r - \bold r^{\prime}| \ll \lambda/\sqrt{\varepsilon (\omega)}$, or if the value of permittivity is in the vicinity of zero: $\varepsilon(\omega) \rightarrow 0$,
the general formula (\ref{Eq5-4}) describes only the longitudinal fields of the oscillating dipoles and the wake field:
\begin{eqnarray}
\displaystyle && \bold E^{pol}\approx \bold E^{stat} = \frac{i}{\omega \varepsilon (\omega )} \int \limits_{V_T} \Big ( 3 \frac{\bold r - \bold r^{\prime}}{|\bold r - \bold r^{\prime}|} \Big (\frac{\bold r - \bold r^{\prime}}{|\bold r - \bold r^{\prime}|}, \bold {j}^{pol (0)}\Big ) - \cr \displaystyle && - \bold {j}^{pol (0)} \Big ) \frac{1}{|\bold r - \bold r^{\prime}|^3}d^3 r^{\prime} 
+ \Big (\frac{1}{\varepsilon (\omega)} - 1 \Big ) \bold E^0 (\bold r, \omega ), \label{Eq5-11}
\end{eqnarray}
and $\bold H^{pol}\approx \bold H^{stat} = 0$. The first term here represents the static field of the dipole moments with volume density $\bold P^{(0)} = \bold j^{(0)}/(-i \omega) = (\varepsilon (\omega) - 1)/(4 \pi) \bold E^0$ induced inside the medium by the particle's field. Of course, the detailed investigation of properties of the longitudinal waves in the medium of a finite conductivity 
(as well as in plasma) must include the consideration of the spatial dispersion of permittivity.

It is this field $\bold E^{stat}$ that acts on the particles being in the wake of a leading particle of a bunch moving according to the law $\bold r = \bold {v}(t) t$ 
through an obstacle with given dielectric properties, or nearby a diffraction radiation target. If the impact parameter in a problem of DR is not too large compared to the wavelength, 
it is possible to consider the particle's energy losses due to the force $\bold F = e \bold E^{stat}$ using directly Eq.(\ref{Eq5-11}) and without the application of reciprocity theorem. 
For example, if a bunch moves through the cylindrical channel with radius $a \precsim \lambda$ in the target of arbitrary conductivity and width (see Fig.\ref{Fig1}), 
the kick due to the wake fields (\ref{Eq5-11}) may be calculated finding the impedances via the standard methods. 
Comparison of the corresponding solution for the special case of the ideally-conducting target with well-known result for a waveguide \cite{Heifets} is of interest, 
but anyway goes beyond the scope of this paper.

\end{document}